# Atomic-scale terahertz near fields for ultrafast tunnelling spectroscopy

V. Jelic[1], S. Adams[1], M. Hassan[1], K. Cleland-Host[1], S. E. Ammerman[1,†],
and T. L. Cocker[1]*

[1]Department of Physics and Astronomy, Michigan State University, East Lansing, MI 48824, USA

[‡]Present address: Empa, Swiss Federal Laboratories for Materials Science and Technology, 8600 Dübendorf, Switzerland

*cockerty@msu.edu

**Lightwave-driven terahertz scanning tunnelling microscopy (THz-STM) is capable of exploring ultrafast dynamics across a wide range of materials with angstrom resolution. In contrast to scanning near-field optical microscopy, where photons scattered by the tip apex are analyzed to access the local dielectric function on the nanoscale, THz-STM uses a strong-field single-cycle terahertz pulse to drive an ultrafast current across a tunnel junction, thereby probing the local density of electronic states. Yet, the terahertz field in a THz-STM junction may also be spectrally modified by the electromagnetic response of the sample. Here, we demonstrate a reliable and self-consistent approach for terahertz near-field waveform acquisition in an atomic tunnel junction that can be generally applied to electrically conductive surfaces. By combining waveform sampling and tailoring with terahertz scanning tunnelling spectroscopy (THz-STS), we comprehensively characterize the tunnel junction and distinguish local sample properties from effects due to terahertz pulse coupling and field enhancement. Through modelling, we verify the presence of an isolated unipolar terahertz-induced current pulse, facilitating straightforward interpretation for differential THz-STS with high spectral resolution. Finally, we demonstrate the feasibility of atomic-scale terahertz time-domain spectroscopy via the extremely localized near-fields in the tunnel junction.**

Lightwave-driven terahertz scanning tunnelling microscopy (THz-STM) uses the near field of a tip-coupled terahertz pulse to coherently control ultrafast electron tunnelling[1,2] and, consequently, resolve terahertz-field-induced rectified charge with atomic resolution[3-8]. Measuring rectified charge as a function of terahertz peak field[5,9] has been established as an analogue of conventional scanning tunnelling spectroscopy, which yields the local density of electronic states (LDOS) on the atomic scale[10-12]. Yet, to accurately extract the LDOS – and, more generally, to understand THz-STM results – the oscillating electric field of the terahertz pulse must be known quantitatively. This presents a challenge, since the frequency response of the tip (which acts as a terahertz wave antenna) substantially alters the amplitude and the phase of the incoupled pulse[4,13-17]. Furthermore, material resonances and terahertz-pulse-induced dynamics in the tunnel junction can alter the nanoscale terahertz near field. The latter effect presents an opportunity: terahertz time-domain spectroscopy[18-20] (THz-TDS) on the atomic scale. However, for this to be practical and versatile, a holistic technique must be developed that both reads out the quantitative local terahertz near field for arbitrary tip position and connects it with the sample LDOS via terahertz scanning tunnelling spectroscopy (THz-STS).

Previously, terahertz fields at the apex of an STM tip have been measured with photoemission sampling[5,14-16] (PES), where an optical pulse incident on the tip excites electrons through multiphoton absorption and the terahertz field modulates the local potential energy landscape[21]. Scanning the relative delay between pulses reveals the terahertz-field temporal profile at the tip apex. While this technique can reliably provide the terahertz near-field waveform across a range of incident optical pulse energies[15], it requires tightly focused pulses with several nanojoules of energy to generate sufficient multiphoton photoemission at the tip apex. When operating at megahertz-scale repetition rates, these pulses deposit a significant heat load onto the tunnel junction and continuously modify the atomic-scale structure of the tip apex, which is critical for atomic-scale imaging and THz-STS. The need for an optical pump pulse also precludes PES as a general approach for local terahertz spectroscopy, since a photoexcited sample complicates the waveform readout. Another method for measuring the terahertz near-field waveform in a THz-STM junction has been to utilize a single-molecule switch[22] to read out the local field[17]. Although this technique is local and non-destructive, it cannot be used at arbitrary tip positions or sample surfaces.

Here, we present an alternative approach in which the local terahertz field is read out via the terahertz-pulse-driven rectified charge ($Q_{THz}$) across the THz-STM junction and validated through self-consistent THz-STS modelling. We introduce terahertz pulse cross-correlations (THz-CC) as a technique for quantitatively and non-destructively measuring terahertz near fields in an atomic tunnel junction. This approach is inspired by reference #17, but utilizes $Q_{THz}$ rather than molecular switching probability and therefore may be applied to a wide variety of conductive surfaces. Our technique uses a large-amplitude terahertz pulse that generates a unipolar subcycle current transient for waveform sampling and a small-amplitude terahertz pulse that is measured. A novel terahertz waveform tailoring scheme enables fine control of this current transient. In a cross-correlation, the weak-field pulse linearly modulates the rectified charge, revealing its waveform. We demonstrate our technique on the reconstructed Au(111)-(22×√3) surface by studying single gold adatoms and comparing them to nearby surface-layer gold atoms that are chemically bonded to the bulk. The spatial distribution of the adatom in THz-STM images is strongly linked to the junction's current-voltage (*I-V*) characteristic and near-field waveform, further emphasizing the need for a comprehensive technique that can reveal both. The gold sample serves three purposes: (i) delicate preparation of the tip apex via controlled nanoscale crashes into the sample surface, (ii) subsequent characterization of the tip's sharpness and radial symmetry for STM and THz-STM imaging of arbitrary samples, and (iii) measurements of the amplitude and phase of the terahertz-pulse-induced voltage within the tunnel junction. Using this concept, we demonstrate the feasibility of terahertz time-domain spectroscopy using a single gold adatom, heralding a new era for terahertz near-field spectroscopy on the atomic scale. Finally, our approach also enables a new differential THz-STS technique with improved sensitivity to features within the LDOS that are otherwise obscured.

## Results

**Subsection 1: Experimental design for lightwave-driven spectroscopy**

In this article, we introduce a toolbox of new techniques that provides a comprehensive experimental description of lightwave-driven tunnelling within an arbitrary atomic tunnel junction. In THz-STM, the temporal profile of the terahertz voltage waveform is an important consideration that has motivated waveform tailoring schemes[13,23,24]. In some cases, a low terahertz transmission through the optics used for waveform tailoring threatens to limit the impact of the scheme. Figure 1a depicts a novel approach with near-unity transmission of the input peak field. Starting with a symmetric (sine-like) terahertz pulse, an optimally asymmetric (cosine-like) pulse can be constructed by equally splitting the original pulse, inverting the polarity of one pulse with respect to the other (I and II in Fig. 1a), and then recombining the fields with a half-period temporal offset (III in Fig. 1a). Notably, the peak field of the constructed asymmetric pulse matches the peak field of the input symmetric pulse, whereas the field at opposite polarity forms two smaller amplitude half-cycles. Figure 1b shows an experimental scheme to realize this scenario. A key aspect of this experimental geometry is a parabolic mirror that focuses the terahertz pulse onto the stationary arm end-mirror of the Michelson interferometer (beamline II in Fig. 1b), thereby inverting the output terahertz field for that arm relative to the other due to the Gouy phase shift[25]. Meanwhile, a stage in the adjustable arm (beamline I in Fig. 1b) is used to precisely control the delay between terahertz pulses ($\tau_\varphi$) and hence the waveform asymmetry. An additional flip mirror, shown in Figure 1c, controls the polarity of the input terahertz field and therefore the polarity of the shaped voltage pulse incident onto the tunnel junction. The polarity inversion can alternatively be performed after waveform shaping.

Although an optimally asymmetric terahertz electric-field pulse can be constructed in free space (and measured coherently with electro-optic sampling, EOS), coupling the terahertz pulse to a metal wire antenna like the STM tip modifies its spectral phase and amplitude across the gap[4,5,13-16], so the waveform asymmetry should be optimized for the near field. In addition to the transfer function of the STM tip, the terahertz voltage transient is also defined by the local dielectric response of the sample, as in near-field microscopy[1,26,27]. Here, we introduce a new technique to measure the terahertz near field, including effects due to the spectral response of the tip and the sample dielectric function, where we utilize our recently established algorithm for analyzing THz-STS measurements[9]. We first separate the asymmetric terahertz pulse into a 'strong-field' pulse, $E_{SF}(t)$, and an opposite polarity 'weak-field' replica, $E_{WF}(t)$, and introduce a delay time between them, $\tau_{CC}$, as shown in Figure 1d. This cross-correlation setup is sketched in Figure 1e, which includes wire-grid polarizers to attenuate the peak field of either pulse without altering the pulse shape[28]. The strong-field terahertz pulse with peak field strength of a few-hundred volts per centimetre is focused onto the STM tip with a parabolic mirror (see Methods), producing $V_{SF}(t)$ across the junction. Meanwhile, the weak-field pulse, with peak field on the few volts per centimetre scale is coupled collinearly to the STM tip and produces $V_{WF}(t)$ across the junction, with $V_{WF}/V_{SF} = -0.03$ (see Extended Data Fig. 1). A focus in the weak-field beamline inverts the polarity of the weak-field pulse relative to the strong-field pulse, while also allowing the weak-field pulse train to be modulated at $f_{CC}$ for differential detection.

When $V_{SF}(t)$ (Fig. 1f, bottom) is tuned to generate a unipolar, sub-half-cycle current pulse across the junction, $i_{SF}(t)$ (Fig. 1f, right), this current pulse can be used to sample the weak-field voltage transient (Fig. 1f, top), similar to photoconductive sampling[29-31] and photoemission sampling[14-16,21]. However, our technique is localized to the atomic tunnel junction used for THz-STM/STS and based entirely on terahertz fields. As visualized in Figure 1f, the weak-field half-cycle that overlaps with the main strong-field peak, at, for example, $\tau_{CC} = \tau_1$, $\tau_2$, or $\tau_3$, modulates the amplitude of the strong-field-induced current pulse (Fig. 1f, right). For sufficiently small $V_{WF,pk}$, the modulated strong-field unipolar current pulse depends linearly on the weak-field amplitude (see Extended Data Fig. 2). Thus, the weak-field voltage transient may be measured (Fig. 1f, top) by scanning $\tau_{CC}$ and recording the change to the rectified charge, $\Delta Q_{THz}$, at the modulation frequency of the weak-field pulse train (Fig. 1g), where $Q_{THz}$ is the temporal integral of the induced current pulse measured in a typical THz-STM experiment (see Methods). Finally, the strength of strong-field voltage pulses across the tunnel junction may be obtained through a calibration based on THz-STS (discussed in the following section), under the assumption that the weak-field and strong-field pulses have the same temporal shape (but opposite polarity). This is justified because the chromium-coated high-resistivity silicon wafers used as terahertz pulse beamsplitters exhibit minimal dispersion within our spectral range[28], as was further confirmed by EOS.

Once a waveform has been measured, it is essential to validate the waveform – i.e.,to confirm that it represents the 'true' terahertz voltage transient at the tunnel junction. This is done by verifying that the strong-field-induced current pulse is unipolar and subcycle; otherwise, physically inaccurate 'false' waveforms are recorded. In the following section, we demonstrate a protocol for this verification and show that we can both validate a true waveform and predict the deviations of a false waveform.

**Subsection 2: Measuring and validating ultrafast oscillating fields in an atomic tunnel junction**

Our procedure for validating a terahertz voltage waveform measured by the cross-correlation sampling technique consists of six phases, as illustrated in Extended Data Figure 3: (i) At a particular location on the sample surface, cross-correlations are measured as a function of $E_{SF,pk}$ (the strength and polarity of the strong-field pulse that drives tunnelling). An example is shown in Figure 2a, with clear differences appearing in the waveform shape at $E_{SF,pk} > 0$ and $E_{SF,pk} < 0$. (ii) The rectified charge is recorded as a function of the peak terahertz electric field, $Q_{THz}(E_{SF,pk})$, which we have previously established as the basis for THz-STS[5,9]. Such a measurement is shown in Figure 2b for the same tunnel junction as Figure 2a. (iii) Following the THz-STS inversion algorithm[9] (with some improvements to the procedure, see Methods), we fit the $Q_{THz}(E_{SF,pk})$ measurement in Figure 2b with a high-order polynomial and extract the differential conductance ($dI/dV$) and current-voltage ($I$-$V$) characteristic (Fig. 2c) acted on by a terahertz pulse with the same field profile as a test waveform (for example, dashed black line in Fig. 2a). (iv) The field-dependent terahertz cross-correlations in Figure 2a are simulated based on the selected test waveform and the extracted $I$-$V$ characteristic, producing a model $\Delta Q_{THz}(\tau_{CC}, V_{SF,pk})$ map (Fig. 2d). (v) To determine the calibration constant $\alpha = V_{SF,pk}/E_{SF,pk}$ for a particular tip apex, we perform $Q_{THz}$-$E_{SF,pk}$ measurements for a set of bias voltages, $V_{d.c.}$ (Fig. 2e). Then, we shift the curves along the x-axis based on the formula $V_{SF,pk} = \alpha E_{SF,pk} + V_{d.c.}$ until the curves all completely overlap (Fig. 2f), thereby determining $\alpha$ = (1 V)/(21 V/cm) for this tip apex and hence the calibrated voltage waveform for both strong-field, $V_{SF}(t)$, and weak-field, $V_{WF}(t)$. The local field enhancement, $F$, is determined from the absolute tip–sample separation, $z_0$, via $F = \alpha/z_0$. (vi) Finally, in Figure 2g the test waveform (dashed black line in Fig. 2a) is compared to the simulated waveform for the corresponding strong-field amplitude (dashed black line in Fig. 2d). The excellent agreement between the measured and simulated waveform shapes confirms that we have indeed measured (and selected as a test) the true voltage waveform across the junction. Equally important, in Figure 2h we show the simulated map of all possible ultrafast current transients produced by the voltage pulse in Figure 2g acting on the I-V characteristic in Figure 2c. The result is indeed unipolar and subcycle for $V_{SF,pk} > 0$ (e.g., dashed black line in Fig. 2h), enabling accurate sampling of the weak-field terahertz waveform with the cross-correlation measurement.

To reinforce the validity of the terahertz voltage waveform in Figure 2g, we take the additional step of predicting the artefacts within a physically inaccurate (false) measured weak-field waveform due to a multi-pulse current transient. Figure 2i shows a false weak-field waveform at $E_{SF,pk} < 0$ (dashed green line in Fig. 2a) and its counterpart from the simulated cross-correlations (dashed green line in Fig. 2d). The simulation reproduces its primary features, providing further confidence that the field profile in Figure 2g is indeed correct. Notably, when the field profile in Figure 2i is used as the initial test waveform, the cross-correlation measurements are not reproduced by the simulation (not shown). The current pulse induced by the strong field that was used to sample the weak-field pulse and obtain the false cross-correlation waveform is shown with a dashed green line in Figure 2h. As expected, the current pulse is multi-cycle and bipolar, and as a result the strong-field-induced current pulses overlap with different parts of the weak-field transient simultaneously, yielding a physically inaccurate waveform. As the asymmetry of the terahertz field (along with the differential conductance) defines the $V_{SF,pk}$ range of unipolar current pulses, waveform shaping is a key technology for accurate cross-correlation sampling with large dynamic range.

**Subsection 3: Tailoring terahertz near-field waveforms in the tunnel junction**

Due to the inherent bipolar nature of electromagnetic transients incident onto the tip, a coherent field-driven process such as THz-STM greatly benefits from precision control of the terahertz pulse phase and field asymmetry. In terahertz cross-correlation measurements, the waveform asymmetry further contributes by making unipolar current pulses more easily accessible. Our experimental approach allows us to design, calibrate and verify a low-loss optimally asymmetric terahertz voltage transient at the tip apex, while simultaneously tunnelling with angstrom resolution.

To explore the local parameter space and identify regions of maximum waveform asymmetry, a two-dimensional map of $Q_{THz}$ as a function of waveform shaping delay ($\tau_\varphi$) and terahertz field strength was acquired (Fig. 3a) based on the experimental concept shown in Fig. 1a–c. We identified that $\tau_\varphi$ = –570 fs (Fig. 3a inset) had desirable characteristics, such as a bipolar $Q_{THz}$ for opposing polarities of $V_{SF,pk}$ and a substantially larger rectified charge compared to other waveform delays. By incorporating a weak-field waveform into the experimental geometry (Fig. 1e), we acquire a corresponding map (Fig. 3b) of the differential rectified charge ($\Delta Q_{THz}$) as a function of the waveform shaping delay and the relative delay between the weak and strong fields ($\tau_{CC}$). Naively, one might consider Figure 3b as a map of all possible terahertz near-field voltage transients in the range –1.25 ps ≤ $\tau_\varphi$ ≤ +1.25 ps; however, testing the validity of each waveform (horizontal cross-section) requires one to perform the analysis summarized in section 2 for every choice of $\tau_\varphi$. Furthermore, the map contains wide ranges of $\tau_\varphi$ where $\Delta Q_{THz}$ is well below the noise floor. This is because the strong-field voltage amplitude is significantly reduced due to deconstructive interference between opposite polarity terahertz fields (also observed in Fig. 3a).

We corroborate our optimally asymmetric near-field terahertz waveform at $\tau_\varphi$ = –570 fs by recording the same parameter space as in Figure 3b with a more conventional technique, namely PES[14-16] (see Methods). By retracting the tip several hundred nanometers and illuminating the apex with a train of 10 nJ ultrafast 515 nm laser pulses, we acquire a PES map of all terahertz near-field voltage transients in the range –1.25 ps ≤ $\tau_\varphi$ ≤ +1.25 ps (Fig. 3c). The THz-CC map (Fig. 3b) and PES map (Fig. 3c) agree reasonably well at negative $\tau_\varphi$, but at positive $\tau_\varphi$ only the two primary half-cycles match-up with the PES map. The discrepancies between the two maps are primarily due to the measurement of false cross-correlation waveforms (as discussed earlier) caused by using multicycle and/or bipolar strong-field-induced current pulses to sample the weak-field waveform. A comparison of the THz-CC and PES waveforms at $\tau_\varphi$ = –570 fs (Fig. 3d) and their spectral amplitude/phase (Fig. 3e) reveals that the two techniques acquire nearly identical waveforms with an asymmetry ratio of 1.5 (the ratio between maximum and minimum half-cycles). Notably, PES with our experimental geometry adds a pre-pulse artefact within the waveform that is not physically present in the true terahertz near field. We interpret this as a reflection of the optical pulse along the tip shank before the arrival of the terahertz pulse.

**Subsection 4: Atomic resolution spectroscopy**

High-resolution scanning tunnelling microscopy is often preceded by perturbative crashes of the tip into a metal surface to restructure the atomic-scale apex. This process is repeated until a desirable tip apex is achieved that can image surfaces with atomic resolution and simultaneously reproduce the expected d$I$/d$V$ spectrum of a known test sample (indicating a relatively featureless tip DOS). The latter is necessary for straightforward analysis when performing tunnelling spectroscopy since all measurements with STM (and hence THz-STM) are intimately coupled to the tip DOS[10-12]. The elegance of our approach is that these tip modifications can be performed as needed, and the calibrated local terahertz near-field voltage can be obtained immediately afterwards. During these tip modifications, we ensure that the tip apex can effectively localize the incident terahertz pulse with sufficient field enhancement, while enduring the extreme tunnel currents regularly seen with THz-STM[4,23,32].

The herringbone reconstruction of the Au(111) surface will occasionally have a single gold adatom located at the corner of a herringbone, as shown in Figure 4a. Notably, the spatial and energy distribution of the adatom electron density differs significantly from that of surface-layer gold atoms that are chemically bonded to the bulk crystal (Fig. 4a inset). Tip shaping was performed with decreasing indentation into the gold surface until a sharp image of a gold adatom was obtained with both STM (Fig. 4b) and THz-STM (Fig. 4c). A constant-tip-height THz-STS imaging dataset (Fig. 4d) shows a negative $Q_{THz}$ over the adatom for $|V_{SF,pk}|$ below the tip and sample work functions (4–5 eV). For $V_{SF,pk} > 0$, $Q_{THz}$ of both the adatom and surface turn positive above the tip work function, though at different $V_{SF,pk}$. The tailored terahertz waveform that was used to acquire the data is shown in Figure 4e (measured over the surface). Terahertz tunnelling spectroscopy ($Q_{THz}$–$V_{SF,pk}$ curves) was performed on and off the adatom (Fig. 4f), and our algorithm yields the extracted differential conductance curves in Figure 4g using the waveform in Figure 4e.

To investigate possible local changes to the terahertz field, we further measured the terahertz voltage waveform on another adatom and nearby on the gold surface using another terahertz waveform (Fig. 4h). The signal-to-noise ratio (SNR) of the waveforms reaches 100–200 after a few minutes of averaging, which allows us to compare subtle differences between the amplitude spectra (Fig. 4i), as in THz-TDS. As the waveforms are very similar, we find here that the differences between gold surface atoms and isolated adatoms observed in THz-STM and THz-STS are a result of changes to the tunnel junction rather than dielectric contrast. Prominently, an increased LDOS below the Fermi level for the adatom leads to the initially negative $Q_{THz}$ over the adatom at positive $V_{SF,pk}$ before the eventual onset of terahertz-pulse-induced field-emission from the tip at higher voltages (Fig. 4f,g). Our measurements nevertheless demonstrate the feasibility of performing THz-TDS on the atomic scale, in addition to THz-STS. This is emphasized in the inset of Figure 4i, where division of the amplitude spectra recorded in the two tip locations shows the flat spectral response of the adatom relative to the surface with exceptional SNR.

**Subsection 5: Differential terahertz scanning tunnelling spectroscopy**

To date, analyses of terahertz tunnelling spectroscopy have depended critically on either modelling[4,5] or our recently introduced inversion algorithm[9]. Here, using the same experimental geometry presented earlier (Fig. 1), we introduce differential THz-STS, which may be directly compared to conventional $dI/dV$ in some cases. In differential THz-STS, the weak-field pulse peak is superimposed on the strong-field pulse peak at $\tau_{CC} = \tau_1$, $\tau_2$, or $\tau_3$ (Fig. 1f) and the amplitude of the strong-field pulse is swept. The measured quantity is the change in rectified charge, $\Delta Q_{THz}$, and the weak-field and strong-field terahertz voltages act in a similar way to the conventional a.c. and d.c. biases used for STS. However, such an analogy is effective only if the strong-field pulse induces a single unipolar current pulse; otherwise, the bipolar nature of the pulse will lead to a complex current profile. While modelling is still necessary to fully interpret all the features within the data, we show here that a measurement of $\Delta Q_{THz}(V_{SF,pk})$ and $\Delta Q_{THz}/Q_{THz}$ can serve directly as a qualitative assessment of the LDOS features accessed by the terahertz pulse.

Field emission resonances (FERs), also known as Gundlach oscillations (Fig. 5a), have been readily observed with STM[10,12,33] and are relatively immune to the wide variety of tip DOS features that often complicate interpretation of other measurements. We utilize FERs on Au(111) as a testing ground for differential THz-STS. First, we identify the FERs of our tip–sample junction with conventional STM $I_{d.c.}$–$V_{d.c.}$ curves acquired on the reconstructed gold surface (Fig. 5b), highlighting the strong nonlinearity near $V_{d.c.} = \pm 4$ V (the onset of field emission) and the appearance of wave-like features at higher absolute voltages. These oscillations become much more pronounced with a differential STS measurement that is normalized by the tunnel conductance via $(dI/dV)/(I/V)$ to reduce the influence of the exponentially increasing current with bias[10-12,34] (orange line in Fig. 5b). Figure 5c shows tip-height-dependent THz-STS measurements in which we directly record the total rectified charge, $Q_{THz}$. The $Q_{THz}$-$V_{SF,pk}$ curves do not show the clear oscillatory features associated with FERs, possibly due to the bipolar and integrating nature of terahertz-pulse-induced charge rectification. Remarkably though, the simultaneously acquired weak-field-modulated measurement of the differential charge, $\Delta Q_{THz}(\tau_{CC} = 0$ ps), contains features that correlate with the observed Gundlach oscillations (Fig. 5d). Inspired by the normalized differential conductance described for conventional STS[34], we normalize the measurements via $|\Delta Q_{THz}/Q_{THz}|$ (Fig. 5e). As a result, the data, which spans multiple decades of $\Delta Q_{THz}$, can be shown on a linear scale with several features emerging that are consistent with Gundlach oscillations. This normalized differential measurement of $|\Delta Q_{THz}/Q_{THz}|$ has higher spectroscopic resolution than THz-STS reported previously[5], providing a clearer view of the phenomenon. Finally, $|\Delta Q_{THz}/Q_{THz}|$ at $V_{SF,pk} < 0$ in Figure 5e agrees best with the normalized differential conductance ($V_{d.c.} < 0$ in Fig. 5b) thanks to an approximately isolated unipolar current pulse for this terahertz polarity (see Extended Data Fig. 4 for the validated waveform).

As a final test, we perform STM (Fig. 5f) and THz-STM (Fig. 5g) distance-voltage spectroscopy ($z$–$V_{d.c.}$), which is a conventional approach (in the STM case) to measuring Gundlach oscillations because it addresses the detection of a small modulation superimposed onto an exponentially increasing background by gradually retracting the tip (inset of Fig. 5f) while maintaining a constant current. The steady-state measurements show clear oscillations, as expected; meanwhile, the THz-STM measurements show similar oscillatory features. Yet, we caution that the d.c. voltage where oscillations appear (Fig. 5g) should not be compared quantitatively without further analysis because of the competing effects of $V_{d.c.}$ and $V_{SF}(t)$.

## Discussion

Comprehensive characterization of the local terahertz near field allows us to examine basic questions regarding the experimental geometry used for coupling pulses to STM tips. We find that the terahertz pulse coupling efficiency (local field enhancement) improves significantly by focusing several hundred micrometers above the tip apex, onto the sharp cusp left by the electrochemical etching process. Conversely, optical pulses used during PES were focused directly onto the tip apex, as verified by bright diffraction peaks visible on a CCD camera actively monitoring the tunnel junction. The tips that were used for the experiments in this study (Extended Data Fig. 5) relied on this improved coupling efficiency to reach the ~10 V peak terahertz amplitude that our tailored pulses could deliver to the tunnel junction. Another observation is that the temporal profile of the near-field terahertz waveform is relatively immune to nanoscale modifications of the tip apex; however, the precise field enhancement (and hence the terahertz voltage calibration) can indeed change, though typically no more than a few tens of percent. Conversely, the macroscopic and mesoscopic tip shape and coupling geometry have a direct influence on the phase of the terahertz field at the tip apex[13-16]. Overall, our approach enables terahertz near-field waveform sampling in any THz-STM junction and will promote comparison and benchmarking across a wide range of instruments, tips and experimental geometries.

Another exciting prospect enabled by the combination of the THz-CC technique with THz-STS inversion is atomic-scale terahertz time-domain spectroscopy. Yet, in order to accurately prescribe small spectral changes to specific sample properties (such as the dielectric function) we must proceed with caution because the measured waveforms are susceptible to artefacts and thus must be validated. This is done by ensuring that the strong-field terahertz pulse drives a single unipolar current pulse with self-consistent modelling. An appropriate current pulse further enables differential THz-STS – where the weak-field modulation acts as a local probe of the energy-dependent differential conductance, similar to the a.c. bias modulation used in conventional STS. The power of this approach was demonstrated on an Au(111) surface through the emergence of Gundlach oscillations in the differential THz-STS measurements; these oscillations were obscured in the conventional THz-STS strong-field sweep. The energy resolution of THz-STS using the strong-field pulse in combination with the inversion algorithm is determined by the polynomial order used to fit the $Q_{THz}$-$E_{SF,pk}$ data. Conversely, we propose that differential THz-STS measurements have their energy resolution set by the weak-field voltage amplitude – an improvement by roughly an order of magnitude to the millivolt scale.

Our THz-CC approach in principle applies to any lightwave-driven tunnelling process and not exclusively those driven by terahertz fields, and therefore should be extendable to higher frequencies[15,35-38], provided the experiments are performed in the strong-field regime. Additionally, pulses of different frequencies may be combined, such as using a strong-field multi-terahertz pulse to sample a weak-field terahertz pulse. It is important to stress that a thorough theoretical analysis based on the THz-STM concept[39-42] will be critical to understand future atomic-scale contrast in terahertz near-field spectroscopy, as an atomic tunnel junction differs significantly from the models typically used to describe scattering-type measurements[1,26,27].

Finally, a key strength of THz-STM is its sensitivity to ultrafast pump-probe material dynamics via the tunnel junction[2,3,7,14,22,43,44]. Meanwhile, scanning near-field optical microscopy at terahertz frequencies has recently advanced from nanoscale THz-TDS to time-resolved terahertz spectroscopy, revealing the ultrafast dynamics of the dielectric function on the 10–100 nm scale[45-49]. We envision the concepts presented here will be similarly extended, enabling time-resolved terahertz spectroscopy to reach the ultimate, atomic scale. Leveraging multi-messenger detection[1,50-52] will provide complementary insight, as the new frontier of ultrafast dynamics on the atomic scale is increasingly unveiled.

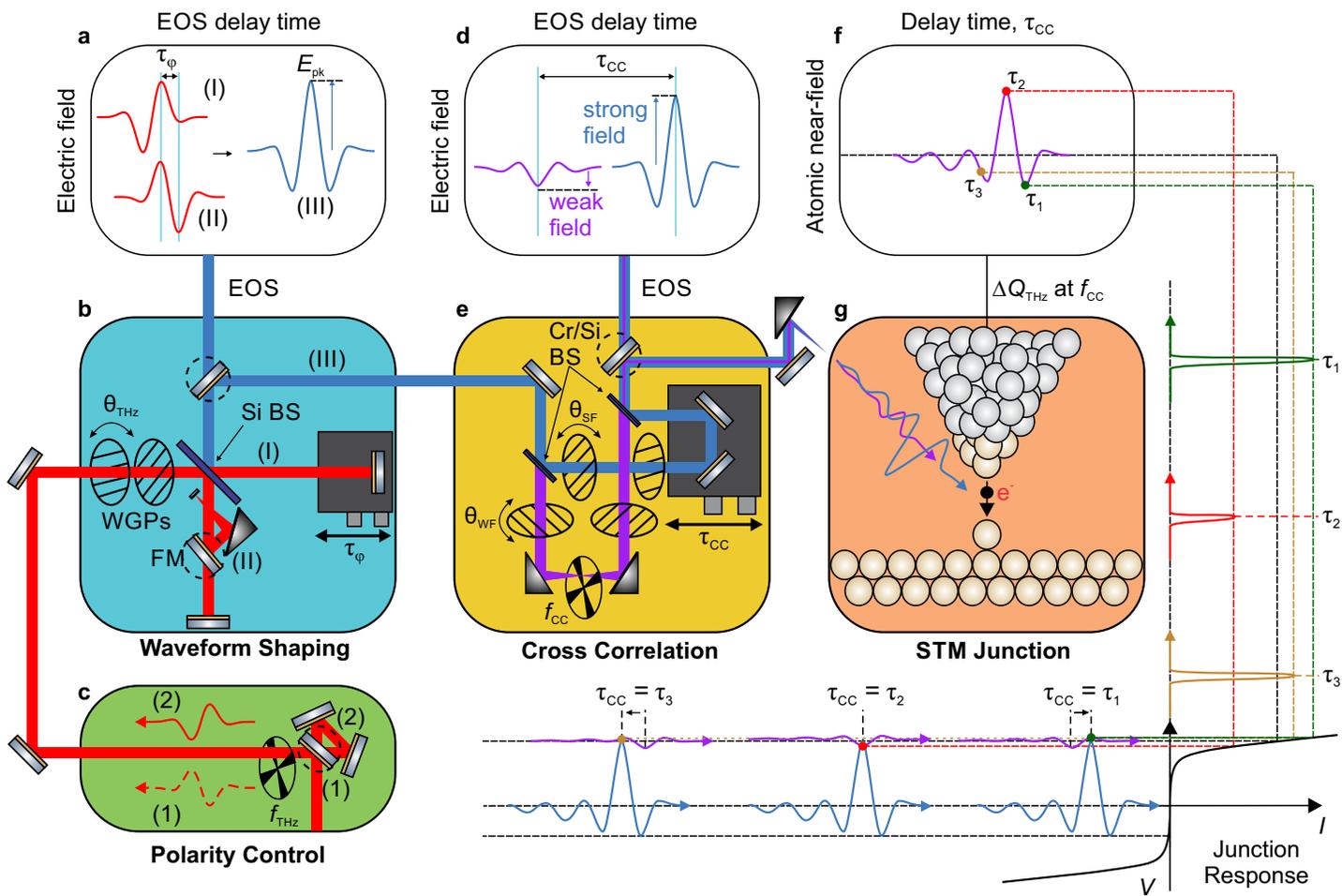

**Fig. 1 | Terahertz waveform shaping and cross-correlation near-field sampling in an atomic tunnel junction. a,** A terahertz (THz) electric-field transient with an asymmetric (cosine-like) waveform shape (III, blue curve) may be constructed through the addition of two symmetric (sine-like) transients of opposite polarity (I and II, red curves) that are offset by a delay time, $\tau_\varphi$, of one-half period. **b,** This configuration is realized experimentally through a Michelson interferometer geometry in which one arm of the interferometer has a parabolic mirror that focuses the THz pulse onto the end-mirror. A flip mirror (FM) enables convenient switching between the waveform shaping setup and terahertz pulse autocorrelations. A silicon beamsplitter (Si BS) provides a spectrally flat response, while a pair of wire grid polarizers (WGPs) enable the peak field strength of the THz pulse to be adjusted via rotation angle $\theta_{THz}$ without affecting the waveform shape. Solid thick red line: original unshaped THz beamline. Solid thick blue line: shaped THz beamline. **c,** Polarity control of the incident terahertz field – and hence the constructed waveform – is achieved with a flip mirror that switches between an odd (1) or even (2) number of reflections. Measurements of the rectified charge, $Q_{THz}$, utilize the optical chopping frequency, $f_{THz}$, for lock-in detection, as does electro-optic sampling (EOS) of the terahertz electric field before it is applied to the STM tip. **d,** For terahertz cross-correlation measurements, the constructed terahertz field is split into a strong-field pulse (blue curve) that drives electron tunnelling across the junction and a weak-field pulse (purple curve) that perturbs the tunnelling probability. **e,** The experimental geometry for cross-correlation measurements is realized through an asymmetric interferometer using two chromium-coated silicon beamsplitters (Si/Cr BS) that enhance the terahertz reflection without affecting the reflected (~83% field) or transmitted (~17% field) waveform shape[28], such that the weak-field pulse (solid thick purple line) is ~3% the strength of the strong-field pulse (solid thick blue line) at the STM junction. WGPs are used to adjust the amplitude of the strong-field THz pulse ($E_{SF,pk}$) and weak-field THz pulse ($E_{WF,pk}$) via rotation angles $\theta_{SF}$ and $\theta_{WF}$, respectively, without affecting the waveform shape. The weak-field pulse train is modulated at an optical chopping frequency of $f_{CC}$ for lock-in detection of the differential rectified charge, $\Delta Q_{THz}$. A delay stage controls the relative arrival time of the strong-field pulse and the weak-field pulse,

$\tau_{CC}$. **f,** Bottom: The combined field for a given $\tau_{CC}$ and $\tau_\varphi$ is strongly enhanced at the STM junction and acts as a voltage on the junction's current-voltage (*I-V*) characteristic. The unipolar current pulse, induced by a carefully tuned (i.e. optimally asymmetric) strong-field waveform (blue curve) samples the weak-field waveform (purple curve) at the corresponding $\tau_{CC}$ (e.g., green pulse: $\tau_1$, red pulse: $\tau_2$, or yellow pulse: $\tau_3$). Top: Scanning the relative delay ($\tau_{CC}$) maps out the weak-field voltage waveform, which has identical shape to the strong-field voltage waveform (though opposite polarity due to focusing through the optical chopper) and corresponds to the field in the atomic tunnel junction. **g,** Schematic of the atomic tunnel junction with the tip positioned over a gold adatom while being illuminated with strong- and weak-field terahertz pulses. The atomic-scale near-field is read out through the differential rectified charge, $\Delta Q_{THz}(\tau_{CC})$ via lock-in detection at $f_{CC}$.

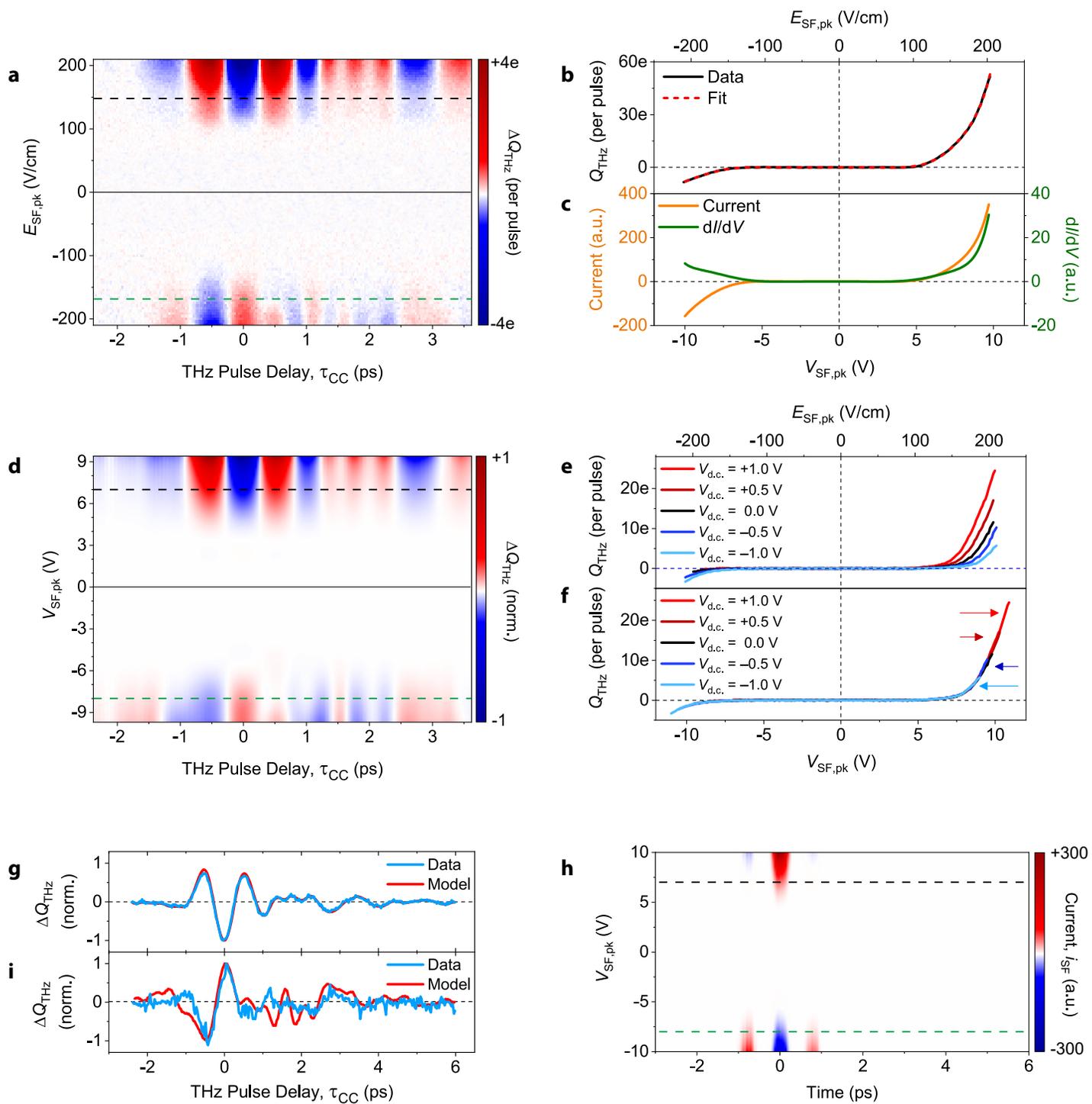

**Fig. 2 | Measuring terahertz near fields in an STM junction. a,** Differential rectified charge map, $\Delta Q_{THz}(\tau_{CC}, E_{SF,pk})$ representing possible terahertz voltage waveforms, measured by modulating the weak-field terahertz pulse (at $f_{CC}$, see Fig. 1) and scanning the delay between the strong-field and weak-field terahertz pulses ($\tau_{CC}$) as well as the strength of the strong-field pulse ($E_{SF,pk}$, via $\theta_{SF}$ in Fig. 1). The data were recorded in constant-current mode with $V_{d.c.}$ = 10 mV and $I_{d.c.}$ = 100 pA. **b,** Rectified charge induced by the strong-field terahertz pulse with increasing incident peak field strength ($Q_{THz}$-$E_{SF,pk}$). The measurement (black curve) was performed at constant height with $V_{d.c.}$ = 0 V, and the tip height set by $V_0$ = 10 mV, $I_0$ = 100 pA. A polynomial fit to $Q_{THz}$-$E_{SF,pk}$ (red dashed curve) acts as input to the terahertz scanning tunnelling spectroscopy inversion algorithm[9], along with a test voltage waveform (dashed black line in **a**). **c,** Extracted differential conductance (d$I$/d$V$, green curve) and extracted current-voltage characteristic ($I$–$V$, orange curve) sampled by the terahertz voltage pulse. **d,** Simulation of $\Delta Q_{THz}(\tau_{CC}, V_{SF,pk})$ based on the extracted $I$-$V$ (**c**), the test waveform temporal profile (dashed black line in **a**), and a weak-field amplitude set to 3% of the strong-field maximum. **e,** $Q_{THz}$-$E_{SF,pk}$ curves as a function of d.c. bias acquired at constant height with the tip retracted by an additional 2 Å from the setpoint $V_0$ = 10 mV, $I_0$ = 100 pA. **f,** Shifted $Q_{THz}$-$E_{SF,pk}$ curves from **e** based on the conversion that 21 V/cm corresponds to 1.0 V ± 0.1 V. The arrows indicate the direction of the applied shift. **g,** Weak-field voltage waveform, $V_{WF}(t)$, across the STM junction (blue curve and dashed black line in **a**). The red curve (dashed black line in **d**) shows the waveform shape determined from the simulation in **d**, confirming that the selected test waveform at $E_{SF,pk}$ = +150 V/cm (dashed black line in **a**) is an accurate representation of the weak-field voltage transient at the tip apex. **h,** Simulated map of current pulses generated by the strong-field voltage waveform applied to the extracted $I$-$V$ characteristic, confirming that a unipolar current pulse was used for the waveform sampling (dashed black line). **i,** THz pulse cross-correlation measurement (blue curve) for $E_{SF,pk}$ = −170 V/cm (dashed green line in **a**). The waveform is notably different from **g**. However, by applying the field profile of the test waveform in **g** to the extracted $I$-$V$ in **c** (with appropriate polarity and field strength, $E_{SF,pk}$ = −170 V/cm), we calculate a matching distorted

waveform (red curve in **i** and dashed green line in **d**), indicating that the misleading shape of the waveform is entirely captured by our model.

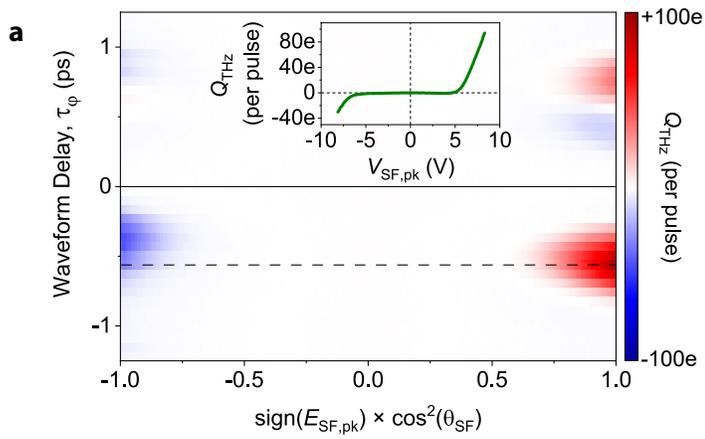
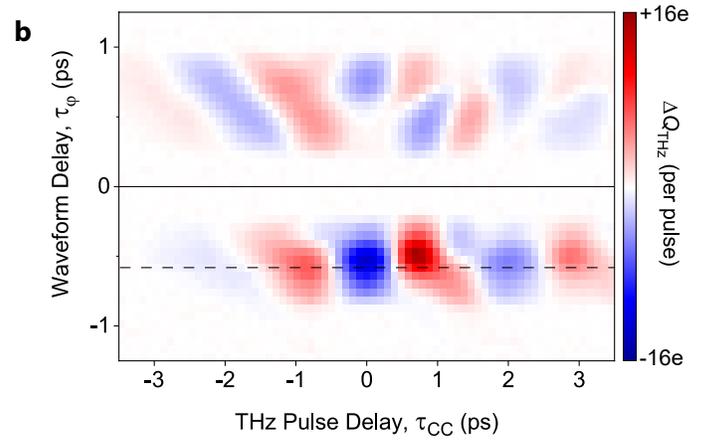
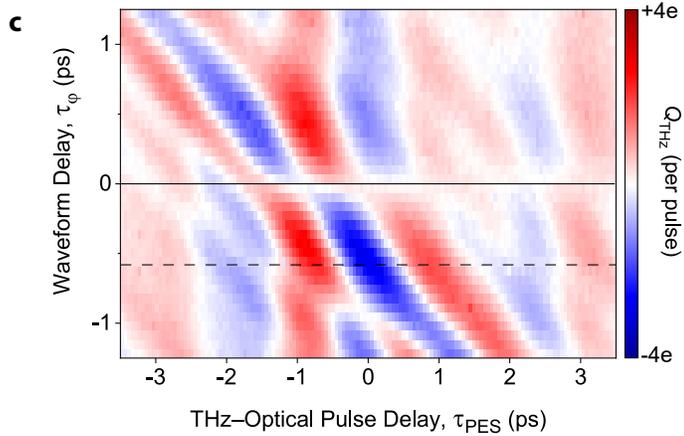
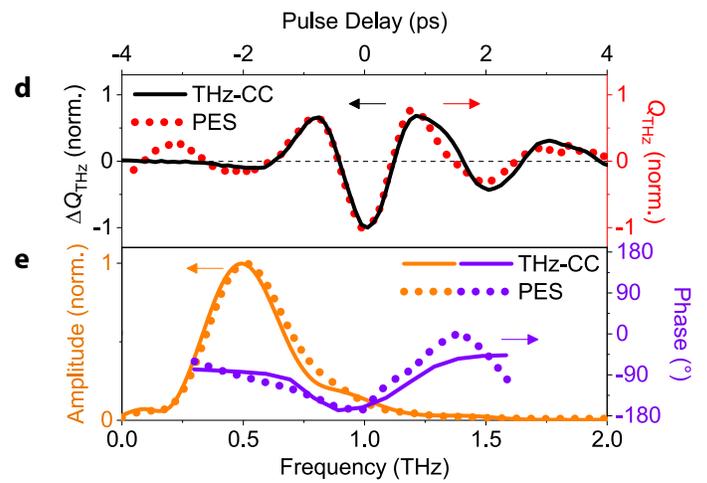

**Fig. 3 | Tailoring terahertz near fields in an atomic-scale tunnel junction. a,** Using the waveform shaping scheme outlined in Fig. 1a,b, a highly asymmetric (cosine-like) terahertz voltage pulse can be identified from a rectified charge map, $Q_{THz}(\theta_{SF},\tau_\varphi)$, for a particular tip and sample. Adjusting the waveform shaping delay, $\tau_\varphi$, modifies the symmetry and peak amplitude of the terahertz near-field waveform, while minor differences in field strength between the two constructing pulses lead to an asymmetry between $\tau_\varphi > 0$ and $\tau_\varphi < 0$. Measurements were performed in constant-current mode with $V_{d.c.}$ = 3 mV, $I_{d.c.}$ = 200 pA. Inset: exemplary $Q_{THz}$–$V_{SF}$ curve at $\tau_\varphi$ = –570 fs (dashed line). **b,** Differential rectified charge map, $\Delta Q_{THz}(\tau_{CC},\tau_\varphi)$, measured at full terahertz electric field strength ($\theta_{SF}$ = 0°). The dashed line indicates $\tau_\varphi$ = –570 fs, where the calibrated strong-field peak voltage is $V_{SF,pk}$ = +8.3 V and the weak-field peak voltage is $V_{WF,pk}$ = –0.25 V. **c,** Terahertz near-field waveforms measured by photoemission sampling (PES) at full terahertz electric field strength ($\theta_{SF}$ = 0°), $V_{d.c.}$ = +6 V and $I_{d.c.}$ = +200 pA with the tip retracted about 300 nm from the Au(111) surface. The terahertz pulse at $\tau_\varphi$ = = –570 (dashed line) has a peak voltage of $V_{SF,pk}$ = +15 V, calibrated with the tip in tunnel contact with the surface via the shift technique presented in Fig. 2e. The optical pump pulse incident onto the tip had a delay of $\tau_{PES}$ with respect to the terahertz pulse, 515 nm center wavelength, ~230 fs pulse duration, 10 nJ pulse energy, ~50 µJ/cm² fluence and a linear polarization along the tip axis (p-polarized). **d,** Comparison of terahertz voltage waveforms measured by terahertz cross-correlation (THz-CC, black curve, dashed line in **b**) and photoemission sampling (PES, red dots, dashed line in **c**). **e,** Spectral amplitude and phase of terahertz voltage waveforms measured by THz-CC (orange and purple curves, respectively) and PES (orange and purple dots, respectively).

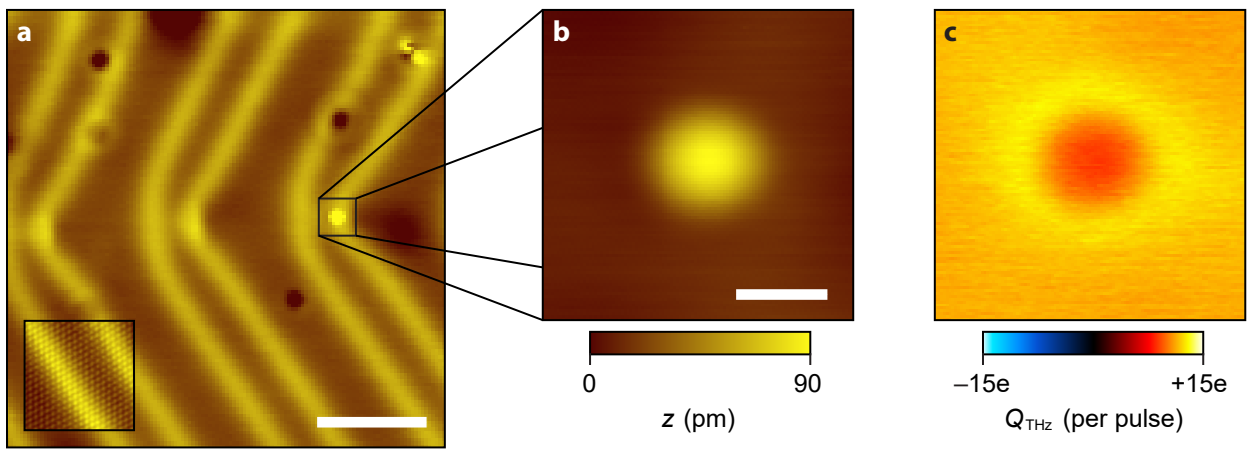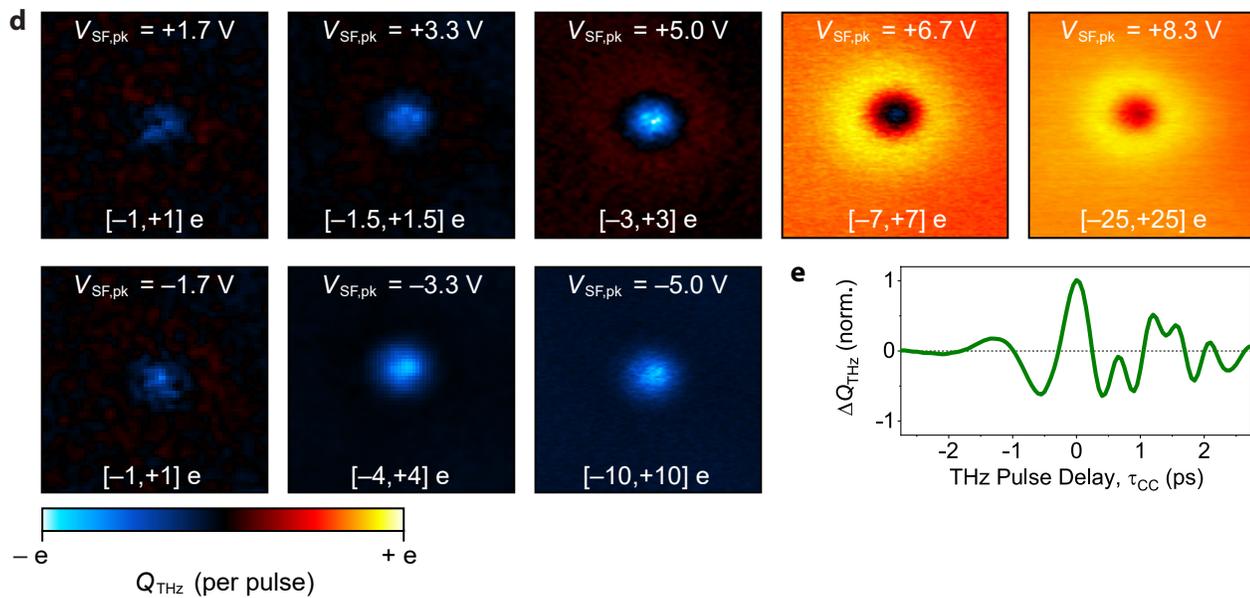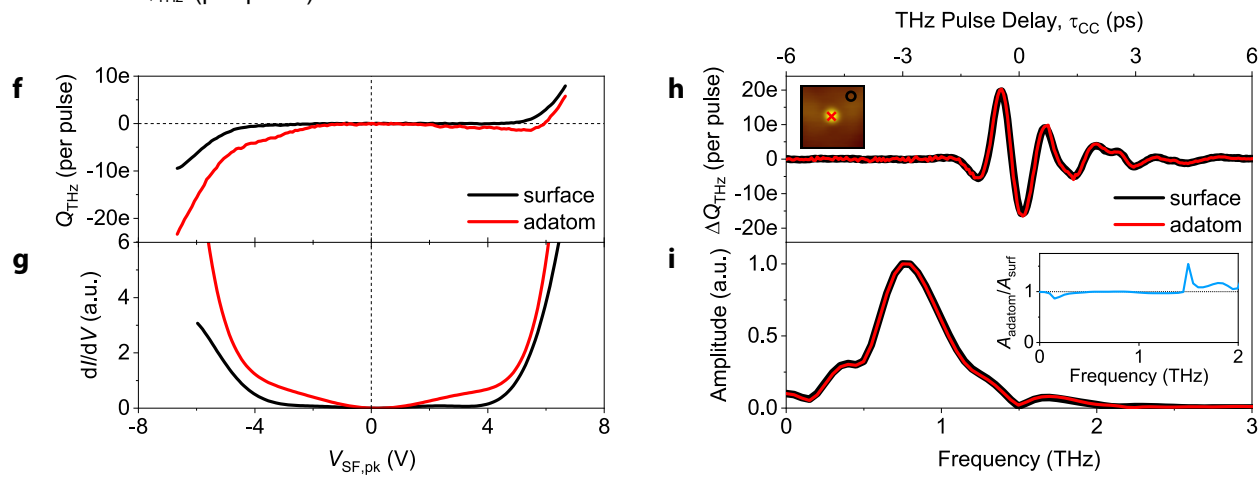

**Fig. 4 | Terahertz scanning tunnelling microscopy and spectroscopy of single gold adatoms.**

**a,** Scanning tunnelling microscopy image of an Au(111) surface recorded in constant-current mode with $V_{d.c.}$ = 1 V, $I_{d.c.}$ = 100 pA. Height range 35 pm; scale bar 5 nm; scan speed 60 nm/s. Inset: high resolution constant-current STM image of the surface with $V_{d.c.}$ = 5 mV and $I_{d.c.}$ = 5 nA, demonstrating the sharpness of the metallic tip (gold-coated tungsten apex). Height range 35 pm; image size 5 nm × 5 nm; scan speed 4 nm/s. **b,** High resolution STM image of a gold adatom on the surface recorded in constant-current mode with $V_{d.c.}$ = 5 mV, $I_{d.c.}$ = 100 pA. Height range 90 pm; scalebar 0.5 nm; scan speed 0.6 nm/s. **c,** Simultaneously recorded THz-STM image of the adatom in **b** with $V_{SF,pk}$ = +8.3 V. **d,** Constant-height THz-STM images of the adatom at different $V_{SF,pk}$. The feedback was disengaged at $V_0$ = 5 mV, $I_0$ = 100 pA, then the tip was approached towards the surface by 50 pm with $V_{d.c.}$ = 0 V. Image size 2 nm × 2 nm; scan speed 0.35 nm/s. **e,** Validated terahertz voltage waveform over the Au(111) surface for the tip apex used to capture **a-d**. **f,** Constant-height $Q_{THz}$–$V_{SF,pk}$ curve for the Au(111) surface and gold adatom shown in **a-d**. The tip height was defined by $V_0$ = 5 mV, $I_0$ = 100 pA over the surface and the tip was approached by 50 pm with $V_{d.c.}$ = 0 V. The constant-height measurements were recorded with the feedback disengaged after the tip had moved to the desired location. **g,** Differential conductance extracted from the terahertz measurements, using the waveform in **e** for both the surface (black line) and adatom (red line). **h,** Another terahertz voltage waveform measured over an adatom and surface at $V_{SF,pk}$ = −9 V. Inset: constant-current STM image showing the waveform measurement locations (red cross: adatom; black circle: surface). Scan parameters: $V_{d.c.}$ = 10 mV, $I_{d.c.}$ = 100 pA; image size 2 nm × 2 nm; colormap range [0,75] pm. **i,** The corresponding amplitude spectra for the surface (black line) and the adatom (red line). Inset: ratio of the two amplitude spectra.

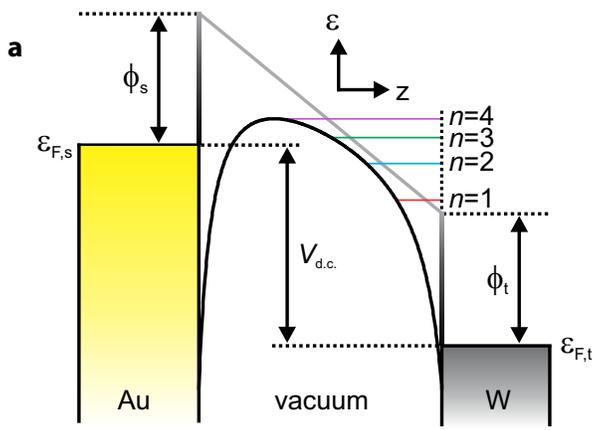
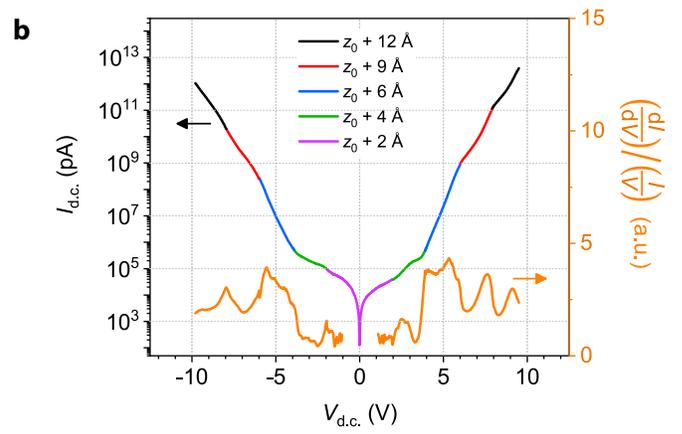
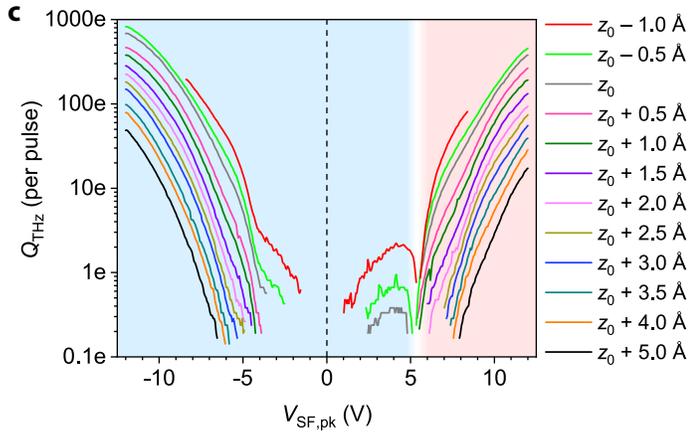
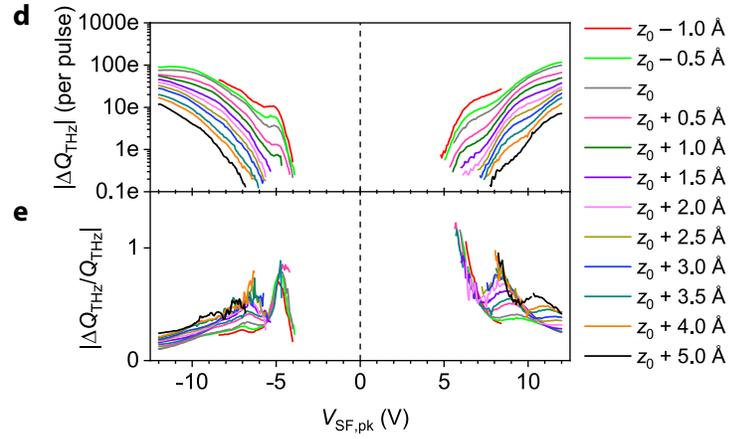
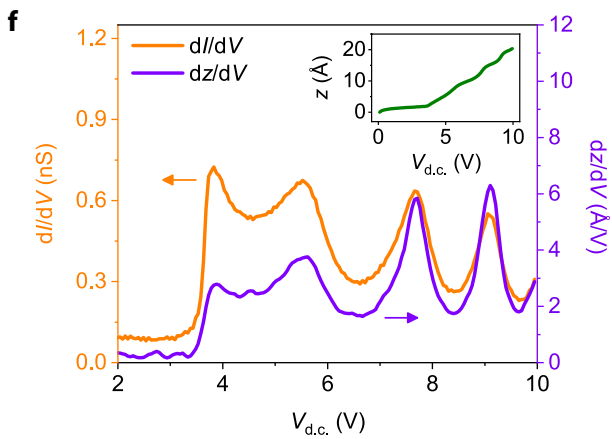
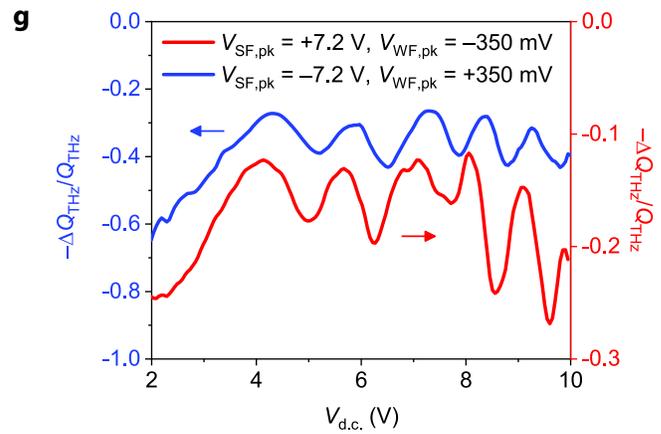

**Fig. 5 | Terahertz tunnelling spectroscopy of Gundlach oscillations on Au(111). a,** Schematic energy–position for a one-dimensional metal-vacuum-metal tunnel junction illustrating the emergence of field emission resonances ($n$ = 1, 2, 3, 4) at a sufficiently high magnitude d.c. bias, beyond the work function of the tip ($\phi_t$) or sample ($\phi_s$). The square barrier approximation is represented by a solid grey line, while the barrier that incorporates image potential effects is represented by the solid black curve. Fermi levels for the tip and sample are given by $\varepsilon_{F,t}$ and $\varepsilon_{F,s}$, respectively, where $eV_{d.c.} = \varepsilon_{F,t} - \varepsilon_{F,s}$. **b,** Constant height current–voltage characteristics ($I_{d.c.}$–$V_{d.c.}$ curves) acquired at several tip–sample separations ($z - z_0$ = 2 Å, 4 Å, 6 Å, 9 Å, and 12 Å for magenta, green, blue, red and black solid lines, respectively) and then normalized to the initial tip height[4]. The initial tip–sample separation, $z_0$, was set by $V_0$ = 10 mV, $I_0$ = 300 pA. The corresponding normalized differential conductance, $(dI/dV)/(I/V)$, is shown with a solid orange line. **c,** Terahertz-pulse-induced rectified charge ($Q_{THz}$) versus peak strong-field terahertz voltage ($V_{SF,pk}$) for the waveform in Extended Data Figure 4g acquired at several different tip heights and $V_{d.c.}$ = 0 V. The initial tip–sample separation, $z_0$, was set by $V_0$ = 2 mV, $I_0$ = 30 pA. Shading indicates the polarity of $Q_{THz}$, with red and blue representing positive and negative, respectively. **d,** Differential rectified charge ($|\Delta Q_{THz}|$) versus peak terahertz voltage ($V_{SF,pk}$) acquired simultaneously with **c**. The weak-field terahertz pulse that was used to modulate the total induced voltage at the tip apex had a peak voltage of $|V_{WF,pk}|$ = 0.35 V and a relative delay of $\tau_{CC}$ = 0 ps with respect to the larger half-cycle of the strong-field pulse. **e,** The corresponding normalized differential rectified charge ($|\Delta Q_{THz}/Q_{THz}|$) versus peak terahertz voltage ($V_{SF,pk}$) for the measurements in **c** and **d**. **f,** Conventional STM constant-current distance–voltage spectroscopy ($z$–$V_{d.c.}$). The derivative of the relative tip height, $z$, with respect to d.c. bias ($dz/dV_{d.c}$) is shown with a solid purple line. The differential conductance ($dI_{d.c}/dV_{d.c.}$) was acquired by applying a 10 mV amplitude a.c. modulation while sweeping $V_{d.c.}$ and retracting the tip height to maintain $I_{d.c}$ = 300 pA (solid orange line). Inset: relative tip height throughout the measurement. **g,** THz-STS distance–voltage spectroscopy ($\Delta Q_{THz}/Q_{THz}$ versus $V_{d.c.}$) acquired at $V_{SF,pk}$ = +7.2 V, $V_{WF,pk}$ = –350 mV, $\tau_{CC}$ = –0.5 ps (red line) and $V_{SF,pk}$ = –7.2 V, $V_{WF,pk}$ = +350 mV, $\tau_{CC}$ = +0.6 ps (blue line), while retracting

the tip to maintain $I_{d.c.}$ = 100 pA. The terahertz waveform used to acquire Figure 5 was not tailored and therefore had a slightly higher amplitude (12 V) and more symmetric shape than the waveform used for Figures 2–4. The validation for this waveform is shown in Extended Data Figure 4.

## Methods

**Terahertz scanning tunnelling microscope.** The STM is a commercial ultrahigh vacuum (UHV) low-temperature STM system with a base pressure of $5 \times 10^{-11}$ mBar and base temperature of 6 K (CreaTec Fischer & Co. GmbH). The base temperature increased to 8 K with the shutter open during THz-STM operation and 12 K during PES with an incident optical pulse train of 10 mW average power. The bias voltage is applied to the sample and the tunnel current is monitored using a preamplifer (Femto DLPCA-200) with a gain of $10^9$ and a bandwidth of 1 kHz. The number of elementary charges rectified by each terahertz pulse is determined via $Q_{THz} = (I_{THz})/(e \times f)$, where f = 1 MHz is the repetition rate, $e$ is the elementary charge and $I_{THz}$ is the measured average terahertz-pulse-induced tunnel current. A ~25-mm-diameter terahertz pulse enters the UHV chamber through a C-cut sapphire viewport and propagates through two additional C-cut sapphire windows on the 77 K and 4.2 K heat shields before being focused onto the tip by a 60° off-axis aluminium parabolic mirror with a focal length of 33.85 mm and a diameter of 25.4 mm. The position of the parabolic mirror is fixed relative to the STM tip. The *p*-polarized terahertz pulses are focused onto the STM tip spanning incident angles of 15°–45° above the horizontal plane (~0.2 sr).

**Sample and tip preparation.** A single-crystal Au(111) sample was cleaned by repeated cycles of argon ion bombardment at 1000 eV and annealing at 850 K. STM tips (Extended Data Fig. 5) were electrochemically etched in a 2 mol/L NaOH aqueous solution from 350 μm diameter polycrystalline tungsten wire and prepared *in situ* with field-directed-sputter-sharpening[53] using 1500 eV argon ions and a +150 V stopping voltage. The tip apex was prepared on the reconstructed gold surface with few-nanometre indentations into the sample surface followed by subsequent apex verification via STM and THz-STM imaging of a gold adatom. The tip indentations were reduced to sub-nanometre values when modifying the last few atoms of the apex and were repeated several times until a sharp THz-STM image of a gold adatom was obtained (Figure 4).

**Terahertz peak voltage calibration.** The peak terahertz voltage was calibrated by acquiring multiple $Q_{THz}$-$E_{SF,pk}$ curves at several $V_{d.c.}$ and then shifting the curves along the x-axis until they all completely overlap as described in the text (see Figure 2e and Extended Data Fig. 4e). The amount of shift required directly determines the local field enhancement factor at the tip apex via $F = \alpha/z_0$, where $z_0$ is the absolute tip height and $\alpha$ is defined by the equation $V_{SF,pk} = \alpha E_{SF,pk} + V_{d.c.}$. We corroborate this voltage calibration in two ways: (i) the THz-STS inversion reproduces our measurements with a value of $V_{SF,pk}$ that agrees with the described terahertz voltage calibration; (ii) a set of conventional and terahertz-pulse-driven approach curves exhibit a similar bias dependence for the extracted apparent barrier heights (Extended Data Fig. 6).

**Qualitative terahertz waveform analysis.** A single, isolated and unipolar terahertz-driven current pulse is highly desirable for reliable near-field waveform acquisition. Accordingly, we have developed a procedure that can be applied during experiments without having to do extensive analysis to verify the measured cross-correlation waveform (though THz-CC across the full range of $V_{SF,pk}$ are not necessary if the initial test waveform is validated). Here, we use terahertz pulse autocorrelations (flip-mirror removed in Fig. 1b) as a diagnostic tool to coarsely indicate if multicycle or bipolar current pulses are present. A typical terahertz pulse autocorrelation at a d.c. bias of 0 V is shown in Extended Data Figure 7a for both positive and negative $V_{SF,pk}$. At negative $V_{SF,pk}$, the rectified charge crosses the line $Q_{THz} = 0$ several times, indicating that a small adjustment of the terahertz pulse phase is enough to completely change the sign of $Q_{THz}$. Therefore, it is not surprising that a waveform acquired at the same settings appears distorted (Extended Data Fig. 7d), suggesting that the strong-field current pulse is bipolar and multicycle. A reduction of the weak-field terahertz amplitude does not affect the waveform distortions (Extended Data Fig. 7d). However, the waveform distortions are not present at either strong-field terahertz polarity when a large d.c. bias voltage is applied (see Extended Data Fig. 7e,f). Autocorrelations acquired under these conditions do not cross the line $Q_{THz} = 0$ (Extended Data Fig. 7b), suggesting that the matching waveforms measured with a large d.c. bias applied were indeed sampled using unipolar current pulses. We consider these waveforms to be promising, but an analysis such as in Figure 2 and Extended Data Figure 4 is still desirable for increased confidence and to ensure isolated current pulses. We further note that applying a large d.c. bias is typically not possible during high-spatial-resolution THz-STM measurements due to the large d.c. background current that would be produced at the tip heights required for THz-STM imaging; therefore, this method is not recommended for atomically resolved studies.

**Photoemission sampling.** The accuracy of measured terahertz cross-correlation waveforms was corroborated with near-field waveform detection at the end of a tip apex using photoemission sampling (Fig. 3 and Extended Data Fig. 8). A 230-fs-duration (full-width-half-maximum) pulse was picked off from the regenerative amplifier laser system and frequency doubled in a β-BBO crystal to a center wavelength of 515 nm before being sent to the STM tip collinear with the incident terahertz pulse. The pulses were set collinear by a 20 Ω/□ indium-tin-oxide-coated BK-7 glass window that was used to reflect the terahertz pulse and transmit the optical pulse. The optical sampling pulse is sufficiently short to omit convolution effects when probing the near-field terahertz waveform[54]. The applied d.c. bias voltage ($V_{d.c.}$) was found to have no effect on the phase of the terahertz field profile measured by PES, as shown in Extended Data Figure 8. This suggests that a large terahertz voltage (as used in our measurements) ignores the waveform distortion effects that would be expected to occur when $V_{d.c.}$ lies within the transition region between linear and exponential scaling in the photoemission *I-V*. Furthermore, we find that our PES voltage calibration[14-16] underestimates the peak terahertz voltage obtained from shifting $Q_{THz}$–$E_{SF,pk}$ curves by more than an order of magnitude (assuming that the induced terahertz voltage at the tip apex does not change with the subwavelength tip retraction during PES[4,8,15]), indicating that more work is needed to understand terahertz waveform measurement and voltage calibration via PES.

**Algorithm for extracting differential conductance from THz-STS.** The steady-state inversion algorithm[9] is used to extract the differential conductance from a THz-STS experiment (excluding a constant offset) if the shape of the near-field waveform in the junction is known. The rectified charge measured in a THz-STS experiment can be described as

$$Q_{\text{THz}}(V_{\text{SF,pk}}) = \int_{-\infty}^{+\infty} I(V_{\text{SF}}(t)) dt, \qquad (1)$$

where $V_{SF}(t) = V_{SF,pk} \times V_0(t)$ is the terahertz-induced near-field waveform with normalized temporal shape $V_0(t)$ and peak voltage $V_{SF,pk}$, which scales linearly with the terahertz field strength.

In the inversion algorithm, we approximate the conductance curve as a polynomial

$$I(V) = \sum_{n=1}^{N} A_n V^n, \qquad (2)$$

where $A_n$ are constant coefficients. We insert this definition of the conductance into Eq. (1) and, after some rearranging, obtain a new expression for the rectified charge

$$Q_{\text{THz}}(V_{\text{SF,pk}}) = \sum_{n=1}^{N} A_n V_{\text{SF,pk}}^n \int_{-\infty}^{+\infty} V_0(t)^n dt \equiv \sum_{n=1}^{N} A_n V_{\text{SF,pk}}^n B_n. \qquad (3)$$

With the terahertz near-field waveform shape $V_0(t)$ characterized, we can easily calculate the $B_n$ terms. The final expression in Eq. (3) is again a polynomial with the only unknown terms being the coefficients $A_n$, which are needed to calculate the I–V curve using Eq. (2).

To find $A_n$, we fit our data $Q_{\text{THz}}(V_{\text{SF,pk}})$ using a polynomial model with order $N$ (Fig. 2b,c: $N$ = 11; Fig. 4f,g: $N$ = 9; Extended Data Fig. 4: $N$ = 14). Note that $A_1$, the linear term, cannot be identified because the integral of the waveform shape $B_1 = 0$. For the fitting procedure we use the Python library scikit-learn[55]. To counteract the problem of underfitting or overfitting, i.e., choosing the right order $N$ for the fit polynomial, we use the shuffle split cross validation method from the same library. Here, we duplicate our original data set multiple times (1000–10,000 times) and split each of these randomly into a training set (80%) and a test set (20%).

For each data set, the polynomial model is trained by performing a fit using the training set. We then use this model to predict the points in our test set and calculate the corresponding mean squared error (MSE). As we vary the model complexity (i.e., the number of polynomial terms) we will find a 'sweet spot' where the MSE reaches a minimum. Going to higher complexity leads to overfitting while decreasing the number of polynomials underfits the underlying true model[56]. Once we have determined the optimal complexity for the fit model, we use the average fit coefficients of all dataset copies to calculate the conductance from Eq. (2).

We use the extracted conductance curve to simulate the rectified charge measurement using Eq. (1) and compare it to the measurement to confirm that the inversion algorithm was performed correctly. In addition, we use the conductance curve to simulate the ultrafast current induced by an incoming terahertz voltage pulse and the terahertz cross correlation measurements.

**Similarities between conventional STS and THz-STS for unipolar current pulses.** As mentioned in the previous section, the THz-STS algorithm enables us to simulate the ultrafast current pulse induced by the terahertz waveform in the tunnel junction. Hence, we can identify certain terahertz peak field strengths ($E_{SF,pk} \propto V_{SF,pk}$), where this current pulse is unipolar. In these cases, the rectified charge can be approximated via

$$Q_{THz} = \int_{-\infty}^{+\infty} I(V_{SF}(t))\mathrm{d}t \propto I(V_{SF,pk}) \int_{-\infty}^{+\infty} \delta(t)\mathrm{d}t. \qquad (4)$$

In addition to measuring the near-field terahertz waveform, we use the weak- and strong-field cross-correlation experimental setup to acquire differential THz-STS measurements, $\Delta Q_{THz}(V_{SF,pk})$ (see Fig. 5). This new measurement approximates the local tunnelling conductance of the sample (d$I$/d$V \propto$ LDOS)[57,58] in unipolar current pulse regions. The mathematical expression for the differential rectified charge is

$$\Delta Q_{THz}(V_{SF,pk}) = \int_{-\infty}^{+\infty} I(V_{SF}(t))\mathrm{d}t - \int_{-\infty}^{+\infty} I(V_{SF}(t) + V_{WF}(t))\mathrm{d}t, \qquad (5)$$

and using the approximation for a unipolar current pulse leads to

$$\Delta Q_{\text{THz}}(V_{\text{SF,pk}}) \propto I(V_{\text{SF,pk}}) \int_{-\infty}^{+\infty} \delta(t) \mathrm{d}t - I(V_{\text{SF,pk}} + V_{\text{WF,pk}}) \int_{-\infty}^{+\infty} \delta(t) \mathrm{d}t \tag{6}$$

$$= I(V_{\text{SF,pk}}) - I(V_{\text{SF,pk}} + V_{\text{WF,pk}}),$$

If $\epsilon = V_{\text{WF}}/V_{\text{SF}} \ll 1$, we can approximate $\Delta Q_{\text{THz}}(V_{\text{SF,pk}}) \propto \mathrm{d}I(V_{\text{SF,pk}})/\mathrm{d}V_{\text{SF,pk}}$ in the case that $V_{\text{SF,pk}}(t)$ induces a unipolar ultrafast current pulse in the tunnel junction.

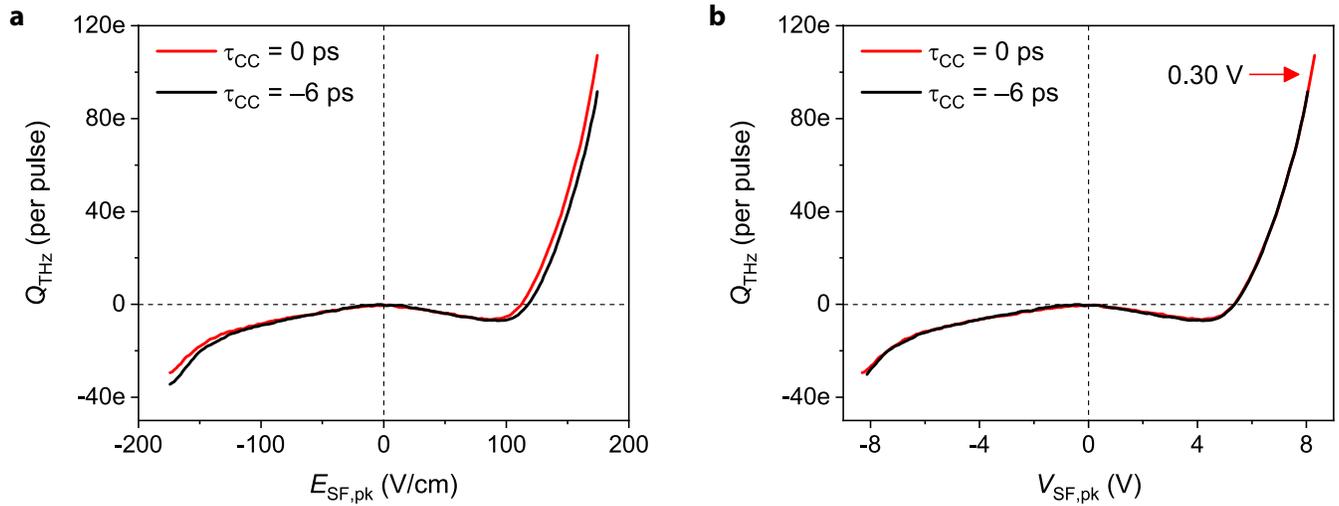

**Extended Data Fig. 1 | Terahertz pulse weak-field voltage calibration. a,b**, Voltage calibration performed for the weak-field waveform, $V_{WF}(t)$, following a similar procedure to Fig. 2f. A pair of $Q_{THz}$–$E_{SF,pk}$ curves (**a**) acquired at $\tau_{CC}$ = 0 ps (solid red line) and $\tau_{CC}$ = –6 ps (solid black line) are used to calibrate the weak-field voltage by translating the red curve along the x-axis by 6.3 V/cm (**b**), equivalent to 0.30 V (3% of $V_{SF,pk}$ = 10 V). The measurements were performed at constant height with $V_{d.c.}$ = 0 V and a 1 Å tip approach from the initial tip–sample separation set by $V_0$ = 2 mV, $I_0$ = 300 pA.

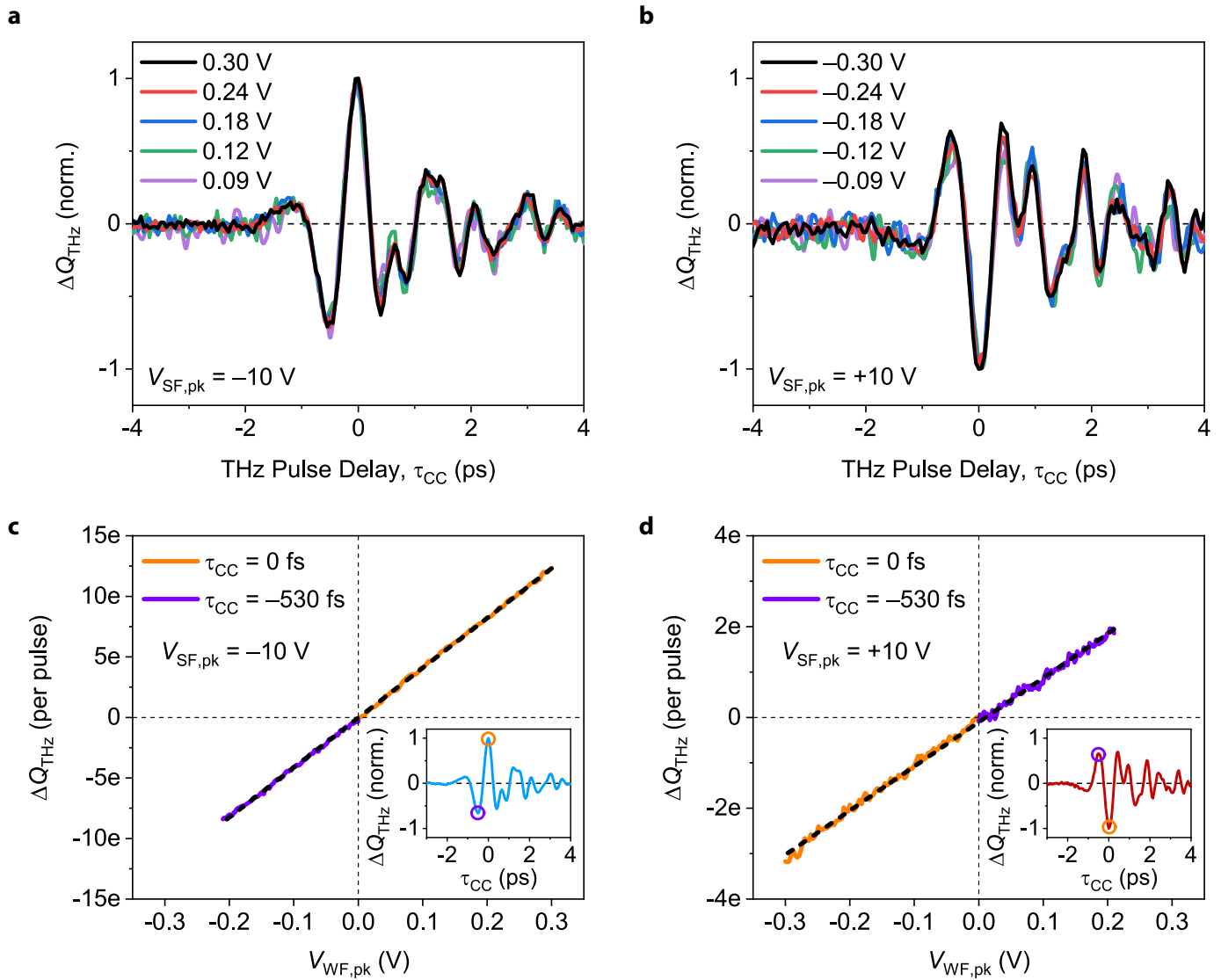

**Extended Data Fig. 2 | Amplitude scaling of the weak-field terahertz pulse. a,b,** Normalized weak-field waveforms, $V_{WF}(t)$, acquired at $V_{SF,pk} = -10$ V (**a**) and $V_{SF,pk} = +10$ V (**b**) for weak-field peak terahertz voltages of $V_{WF,pk} = \pm 0.30$ V (black line), $V_{WF,pk} = \pm 0.24$ V (red line), $V_{WF,pk} = \pm 0.18$ V (blue line), $V_{WF,pk} = \pm 0.12$ V (green line), and $V_{WF,pk} = \pm 0.09$ V (purple line). The measurements were performed at constant current with $V_{d.c.} = 10$ mV, $I_{d.c.} = 200$ pA. **c,d,** $\Delta Q_{THz}$–$V_{WF,pk}$ curves acquired at $\tau_{CC} = 0$ fs (orange line) and $\tau_{CC} = -530$ fs (purple line) for a negative-dominant terahertz polarity at $V_{SF,pk} = -10$ V (**c**) and a positive-dominant terahertz polarity at $V_{SF,pk} = +10$ V (**d**). The insets within **c** and **d** show the corresponding verified weak-field terahertz waveforms measured with THz-CC. The weak-field waveform has an inverted polarity with respect to the strong-field waveform (not shown) due to the experimental geometry (see Fig. 1e). The orange and purple circles within the insets indicate the temporal locations of the corresponding measurements. The measurements were performed at constant height with $V_{d.c.} = 0$ V and the tip–sample separation set by $V_0 = 10$ mV, $I_0 = 300$ pA. The dashed black lines are linear fits to the data.

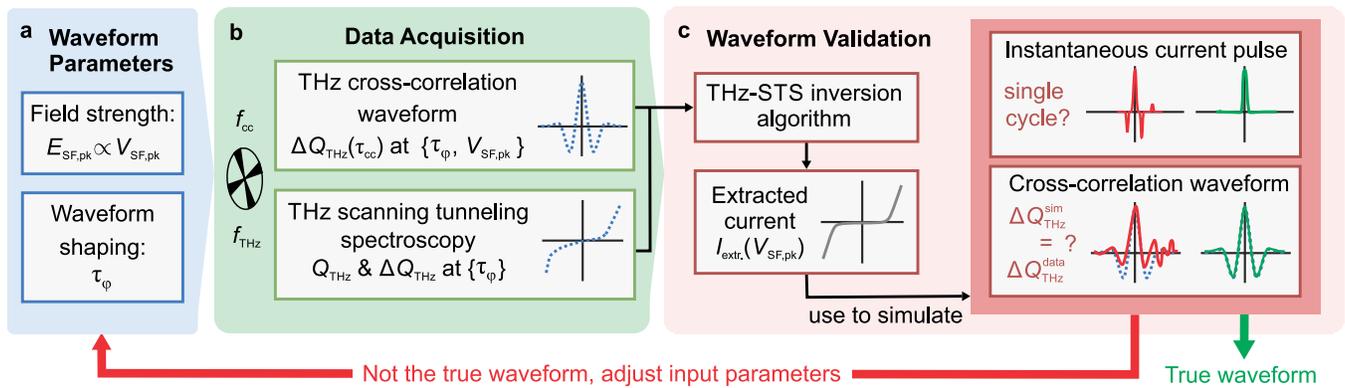

**Extended Data Fig. 3 | Validating terahertz near fields in an atomic tunnel junction.**
**a**, Asymmetric terahertz pulses with an incident field strength of $E_{SF,pk}$ and near-field voltage of $V_{SF,pk}$ are constructed by varying the terahertz pulse interferometer delay time ($\tau_\varphi$) after inverting the polarity of one interferometer arm with respect to the other. **b**, The measurements required for waveform validation at a particular $\tau_\varphi$ are a THz-CC field profile ($\Delta Q_{THz}$ as a function of $\tau_{CC}$) and a $Q_{THz}$–$V_{SF,pk}$ curve. **c**, The THz-STS inversion algorithm uses both measurements to extract the current–voltage characteristic for the tunnel junction that is used to re-rectify the measured terahertz field profile. This produces a simulated current pulse and simulated cross-correlation waveform. If the simulated current pulse is unipolar and the measured and the simulated cross-correlation waveforms match, then the measured field profile is consistent with the model and verified to be accurate.

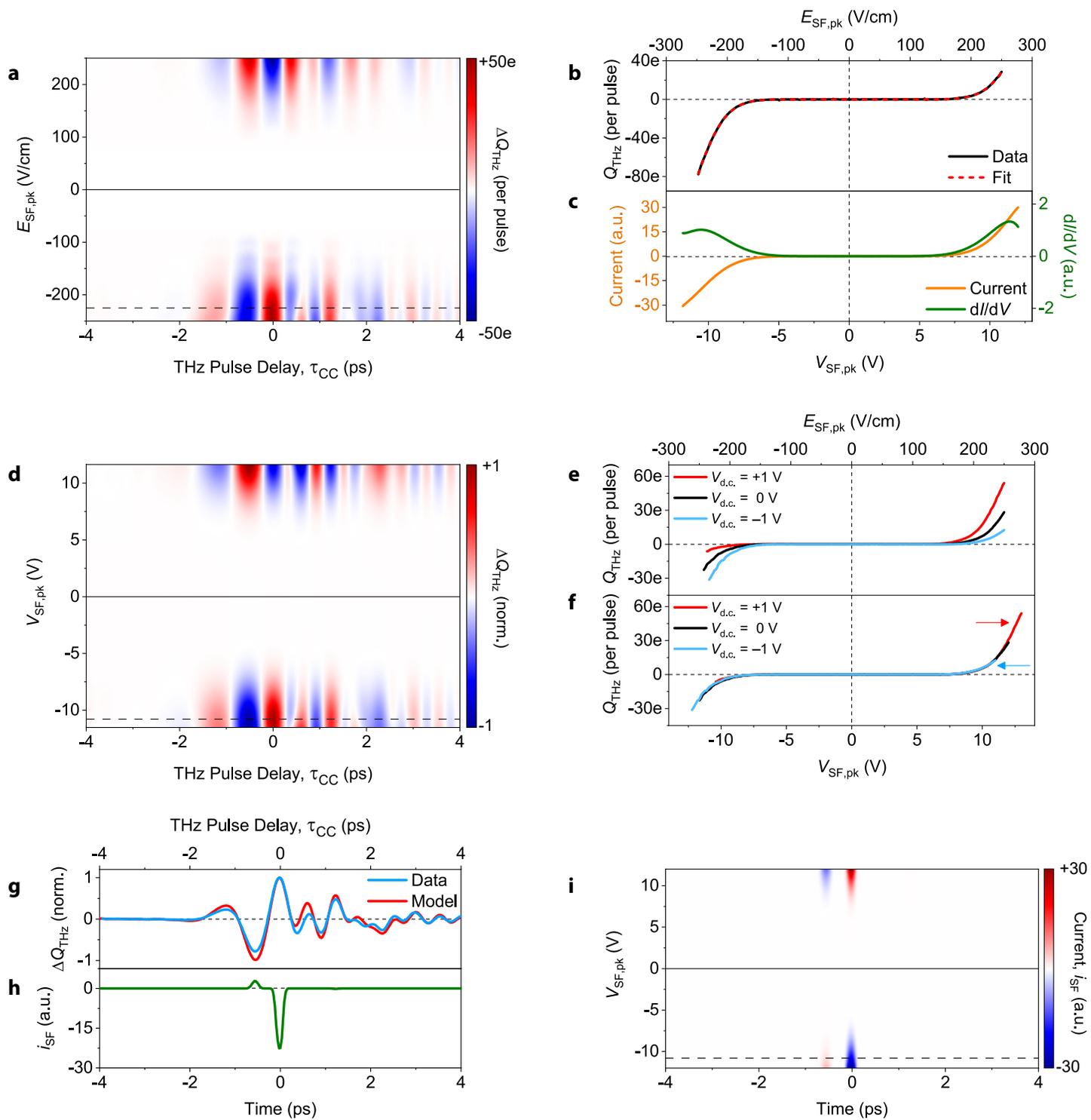

**Extended Data Fig. 4 | Measuring non-tailored terahertz near-fields in an STM junction.**
**a**, Differential rectified charge map, $\Delta Q_{THz}(\tau_{CC}, E_{SF,pk})$ measured by scanning the delay between the strong-field and weak-field terahertz pulses ($\tau_{CC}$) as well as the strength of the strong-field pulse ($E_{SF,pk}$) in constant-current mode with $V_{d.c.}$ = 10 mV and $I_{d.c.}$ = 100 pA. **b**, Measured $Q_{THz}$-$E_{SF,pk}$ (solid black curve) acquired with constant height at the tip height set in **a** and $V_{d.c.}$ = 0 V. A polynomial fit to $Q_{THz}$-$E_{SF,pk}$ (dashed red curve) acts as input to the THz-STS inversion algorithm, along with a test voltage waveform (dashed black line in **a**). **c**, Extracted differential conductance (d$I$/d$V$, green curve) and extracted current-voltage characteristic ($I$–$V$, orange curve) sampled by $V_{SF}(t)$. **d,** Simulation of $\Delta Q_{THz}(\tau_{CC}, V_{SF,pk})$ based on the extracted $I$-$V$ (**c**), test waveform temporal profile (dashed black line in **a**), and a weak-field amplitude set to 3% of the strong-field maximum. **e,** $Q_{THz}$-$E_{SF,pk}$ curves versus $V_{d.c.}$ acquired at constant height with the tip retracted by an additional 4 Å from the setpoint $V_0$ = 10 mV, $I_0$ = 300 pA. **f,** Shifted $Q_{THz}$-$E_{SF,pk}$ curves from **e** based on the conversion that a 21 V/cm terahertz peak field at the tip apex corresponds to 1.0 V ± 0.1 V at the tunnel junction. The arrows indicate the direction of the applied shift. **g,** Weak-field voltage waveform, $V_{WF}(t)$, across the STM junction (blue curve and dashed black line in **a**). The red curve (dashed black line in **d**) shows the waveform shape from **a** applied to the $I$–$V$ curve in **c**, confirming that the selected test waveform at $E_{SF,pk}$ = −225 V/cm (dashed black line in **a**) is an accurate representation of the weak-field voltage transient at the tip apex. **h,** Simulated current pulse generated by the strong-field voltage waveform (with the field profile in **g**) applied to the extracted $I$-$V$ characteristic. **i,** Simulated map showing all possible strong-field-induced current pulses in the range of $V_{SF,pk}$ that was explored. The non-tailored pulse (as-is from LiNbO$_3$) struggles to produce a single unipolar current pulse and hence requires modelling for accurate assessment of the terahertz-induced voltage pulse. A single unipolar current pulse can be imposed by setting a large d.c. bias and retracting the tip to ensure a low d.c. current (see Extended Data Fig. 7), but this setting is not compatible with high resolution THz-STM imaging.

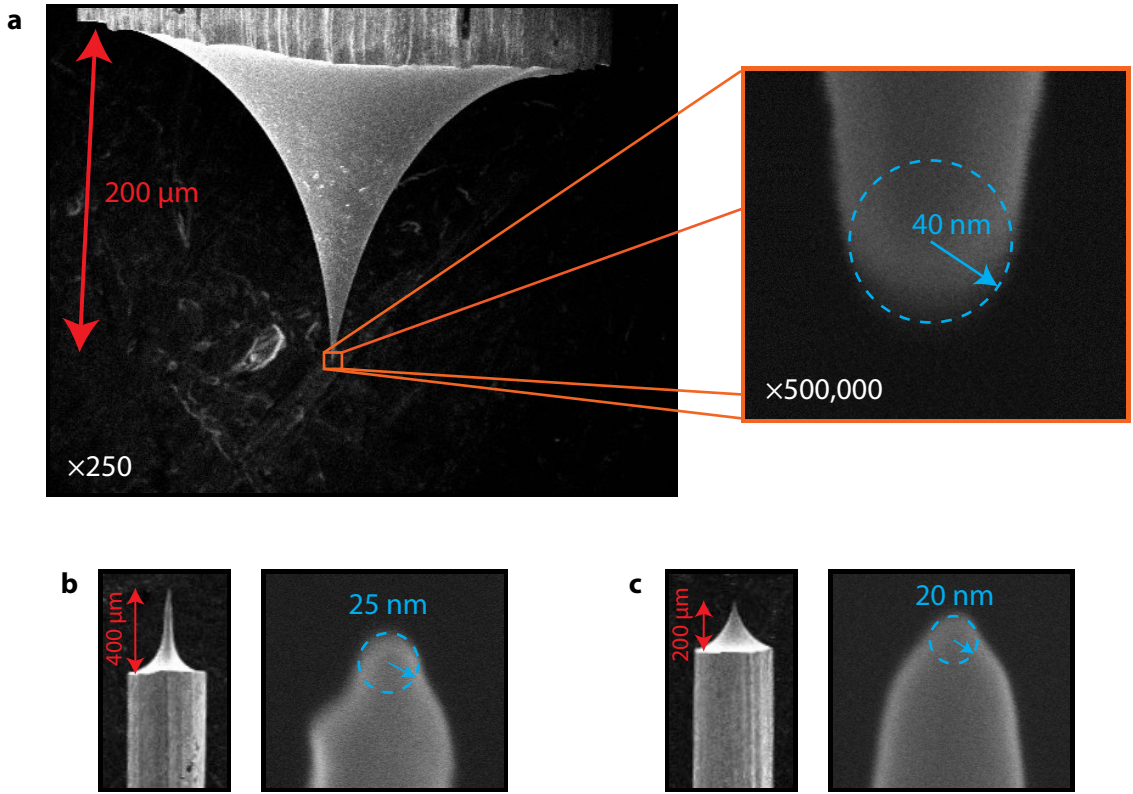

**Extended Data Fig. 5 | Scanning electron microscope (SEM) images of THz-STM tips. a–c,** SEM images of freshly etched polycrystalline tungsten tips at a magnification of ×250 (left) and ×500,000 (right). The images were acquired using the in-lens detector with a 15 keV electron beam propagating at 90° with respect to the tip axis. The red arrows denote cusps with heights of 200 µm (**a**), 400 µm (**b**) and 200 µm (**c**), while the blue circle indicates a radius of curvature at the tip apex of 40 nm (**a**), 25 nm (**b**) and 20 nm (**c**). The tip in **a** was used for most of the measurements. All three tips effectively coupled terahertz pulses to the atomic-scale tunnel junction.

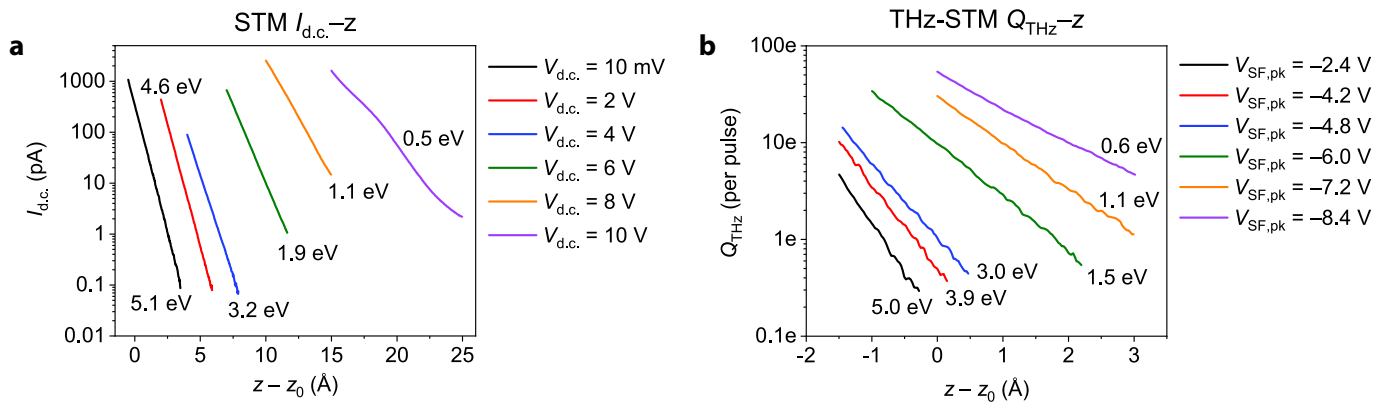

**Extended Data Fig. 6 | Tunnel gap spectroscopy on Au(111) with STM and THz-STM. a,b**, Conventional (**a**) and terahertz-pulse-driven (**b**) approach curves acquired with the feedback loop disengaged at the tip height, $z_0$, set by $V_0 = 10$ mV, $I_0 = 300$ pA. The apparent barrier height is calculated using $\phi = \hbar^2/(8m)\cdot(d\ln(I)/dz)^2$ following an exponential fit (not shown), where $m$ is the electron mass, $\hbar$ is the reduced Planck constant, and $I$ is the d.c. current (**a**) or rectified charge (**b**). The value of $\phi$ in electronvolts is shown beside each measurement. The measurements in **a** were acquired with $V_{SF,pk} = 0$ V, while the measurements in **b** were acquired with $V_{d.c.} = 0$ V.

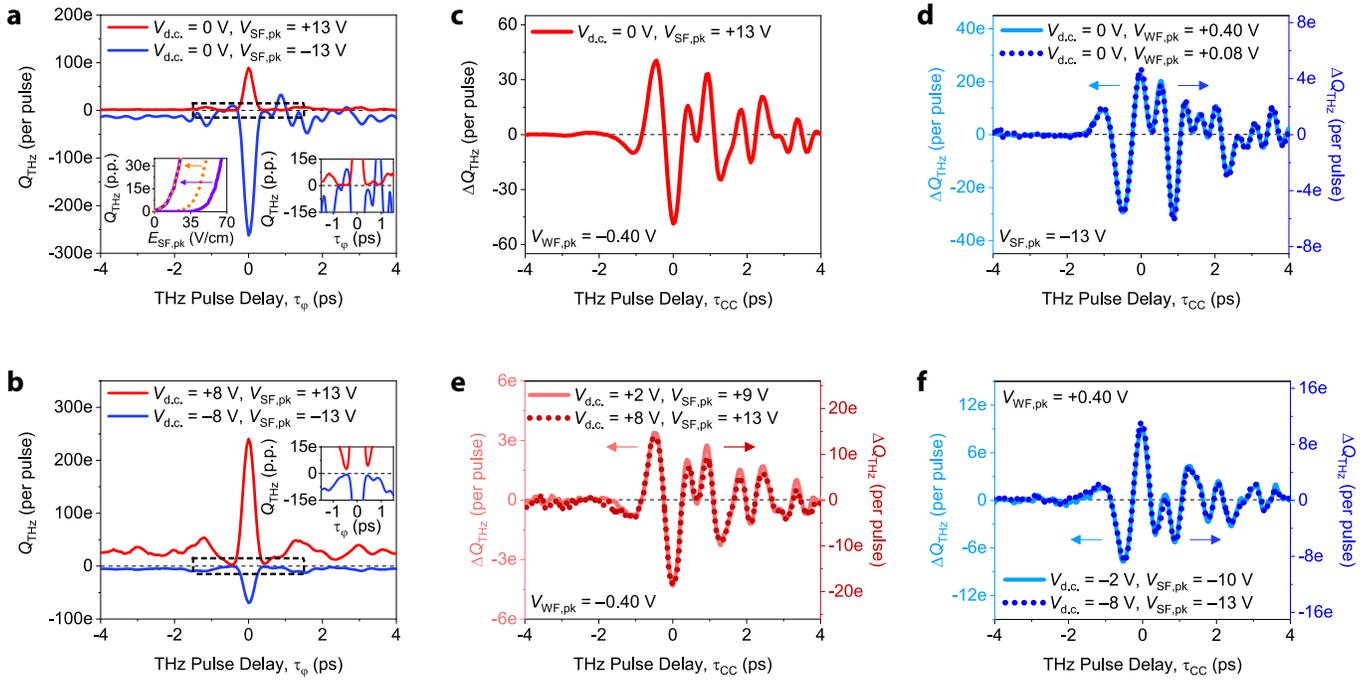

**Extended Data Fig. 7 | Imposing a unipolar strong-field-induced current pulse with a large d.c. bias voltage offset and qualitative waveform verification using terahertz pulse autocorrelations. a,b,** Terahertz pulse autocorrelations at $V_{SF,pk} = +13$ V (solid red line) and $V_{SF,pk} = -13$ V (solid blue line) with $V_{d.c.} = 0$ V, $z = z_0$ (**a**) and $V_{d.c.} = \pm 8$ V, $z = z_0 + 9$ Å (**b**). The left inset in **a** shows a set of shifted and unshifted $Q_{THz}$-$E_{SF,pk}$ curves for the voltage calibration with this tip apex, acquired at $V_{d.c.} = +5$ V (orange circles) and $V_{d.c.} = +8$ V (solid purple line) and a tip height of $z = z_0 + 12$ Å. The right insets in **a** and **b** are a zoom into the central region of the autocorrelation (dashed black line). **c,** THz-CC waveform measured at $V_{d.c.} = 0$ V, $V_{SF,pk} = +13$ V and $V_{WF,pk} = -0.40$ V. **d,** THz-CC waveforms measured at $V_{WF,pk} = +0.40$ V (light blue line) and $V_{WF,pk} = +0.08$ V (dark blue circles) with $V_{d.c.} = 0$ V and $V_{SF,pk} = +13$ V. **e,f,** THz-CC waveforms measured with a large d.c. bias of $V_{d.c.} = \pm 2$ V (red and blue lines) and $V_{d.c.} = \pm 8$ V (red and blue circles) applied during the measurement. The sign of the d.c. bias was matched to the sign of the strong-field terahertz amplitude: +9 V (light red line), +13 V (dark red circles), −10 V (light blue line) and −13 V (dark blue circles). The weak-field terahertz amplitude was −0.40 V for **e** and +0.40 V for **f**. All measurements were performed at constant height with the feedback disengaged at $V_0 = 20$ mV, $I_0 = 200$ pA.

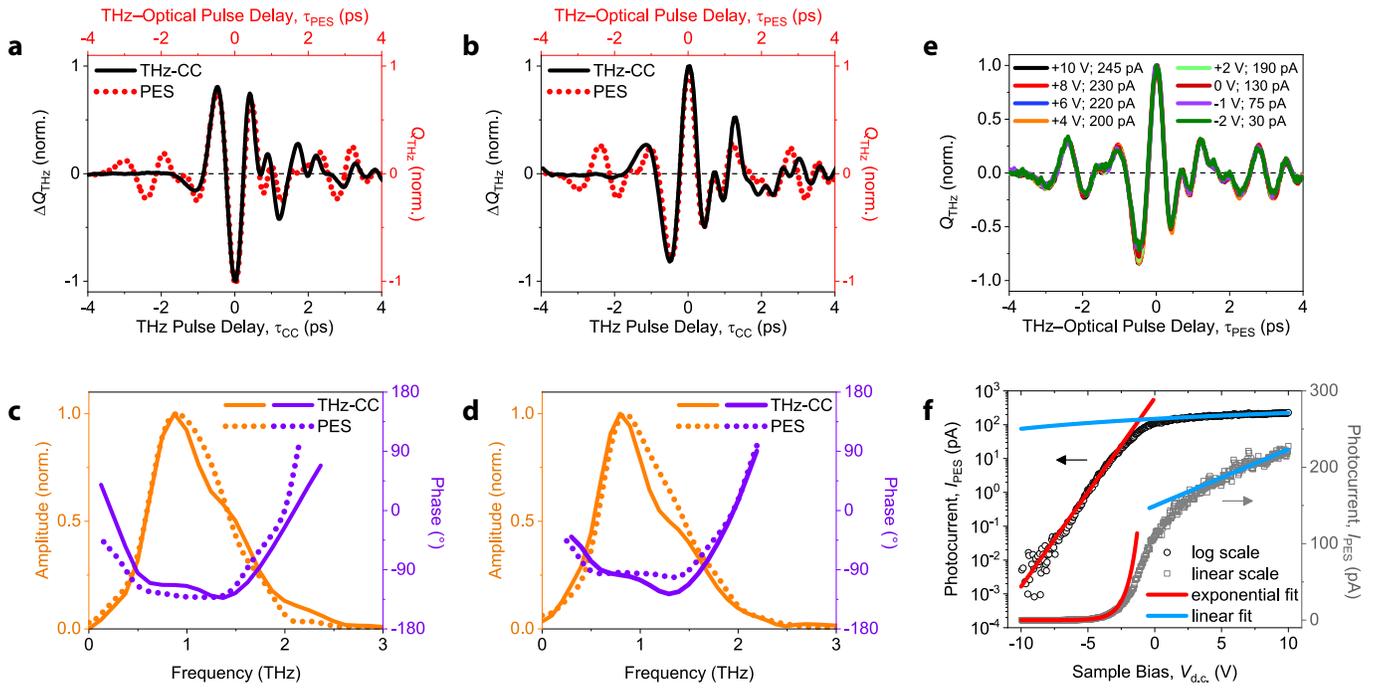

**Extended Data Fig. 8 | Photoemission sampling (PES) voltage dependence and comparison to terahertz cross-correlation (THz-CC) waveforms. a,b**, Measured THz-CC waveforms (solid black lines) acquired at $V_{d.c.} = 0$ V, $V_{SF,pk} = +9$ V (**a**) and $V_{d.c.} = 0$ V, $V_{SF,pk} = -9$ V (**b**) with the feedback loop disengaged at $V_0 = 10$ mV, $I_0 = 300$ pA. The PES waveforms (red circles) were acquired with the tip retracted several hundred nanometers from the Au(111) surface at $V_{d.c.} = +6$ V, $V_{SF,pk} = -15$ V (**a**) and $V_{d.c.} = +6$ V, $V_{SF,pk} = +15$ V (**b**). The *p*-polarized optical pulses used for PES had a 515 nm center wavelength, ~230 fs duration, ~50 µJ/cm$^2$ fluence on the tip apex and 10 nJ pulse energy (accounting for losses through the three sapphire windows). The terahertz voltage calibration for all waveforms was determined from the $Q_{THz}$-$E_{SF,pk}$ curves in Extended Data Figure 4e. **c,d**, The corresponding spectral amplitude (orange) and spectral phase (purple) for the THz-CC (solid lines) and PES (dotted circles) waveforms in **a** (**c**) and **b** (**d**). A window function was used to suppress the prepulse artefact in PES at –2.5 ps for the fast Fourier transforms. **e**, Normalized PES waveforms at multiple $V_{d.c.}$ with the same terahertz and optical pulses as in **b**. The settings within the legend denote the applied d.c. bias to the sample ($V_{d.c.}$) and the resulting total d.c. current ($I_{PES}$). Notably, the terahertz field profile does not change with $V_{d.c.}$ ranging from –2 V to +10 V. The field profile at increasingly negative voltages also looked similar, but the waveforms became noisy (not shown). **f**, Photoemission current–voltage characteristic shown with both log-scale (black circles) and linear-scale (grey squares) used for the y-axis. An exponential fit (solid red line) and linear fit (solid blue line) highlight two distinct regimes within the curve. The terahertz field was set to zero ($V_{SF,pk} = 0$ V) and the optical pulses had the same conditions as in **a** and **b**.

## Data availability

The data that support the plots within this paper and other findings of this study are available from the corresponding author upon reasonable request.

## Code availability

The algorithm for subcycle terahertz scanning tunnelling spectroscopy and the simulation of terahertz pulse cross-correlation measurements were implemented using custom-made Python functions, which will be made open-source upon journal publication of this work. The code will be available on the Github repository (https://github.com/NanoTHzCoding/THz_STS_Algorithm). The codes and simulation files that support the plots and data analysis within this paper are available from the corresponding author on reasonable request.

## Acknowledgements

The authors thank R. Loloee and R. Bennett for technical support and B. Schuler for valuable discussions. This project was financially supported by the Office of Naval Research (Grants no. N00014-21-2537, N00014-21-1-2682), the Army Research Office (Grant no. W911NF2110153), the Air Force Office of Scientific Research (Grant no. FA9550-22-1-0547), and the Cowen Family Endowment.

## Author contributions

V.J., K.C.-H., M.H. and T.L.C. designed and constructed the experimental setup. The experiments were carried out by V.J., M.H., S.A. and S.E.A. with support from K.C.-H. and T.L.C. The modeling was performed by S.A. with support from S.E.A., M.H., V.J. and T.L.C. M.H., S.E.A and V.J. prepared the samples and tips. V.J., S.A., M.H. and T.L.C. wrote the manuscript with contributions from all authors. T.L.C. supervised the project.

## Competing interests

Authors declare no competing interests.

## Additional information

**Correspondence and requests for materials** should be addressed to T.L.C.